\documentclass[amsmath,amssymb,nofootinbib,notitlepage,superscriptaddress,twocolumn]{revtex4-1}
\pdfoutput=1

\usepackage{aas_macros,esint,graphicx,overpic,xcolor}
\usepackage[bottom]{footmisc}
\usepackage[colorlinks=true]{hyperref}
\definecolor{red}{RGB}{228,26,28}
\definecolor{green}{RGB}{77,175,74}
\definecolor{blue}{RGB}{55,126,184}
\definecolor{purple}{RGB}{152,78,163}

\let\originalleft\left
\let\originalright\right
\renewcommand{\left}{\mathopen{}\mathclose\bgroup\originalleft}
\renewcommand{\right}{\aftergroup\egroup\originalright}

\newcommand{\ab}[1]{\left|#1\right|}
\newcommand{\br}[1]{\left[#1\right]}
\newcommand{\cu}[1]{\left\{#1\right\}}
\newcommand{\pa}[1]{\left(#1\right)}
\newcommand{\ed}{\mathop{}\!\mathrm{d}}
\newcommand{\pd}{\mathop{}\!\partial}
\renewcommand{\O}[1]{\mathcal{O}\pa{#1}}

\newcommand{\rt}{\tilde{r}}
\newcommand{\Rt}{\tilde{R}}
\DeclareMathOperator\sign{sign}
\DeclareMathOperator\sn{sn}

\begin{document}

\title{Photon Emission from Circular Equatorial Kerr Orbiters}

\author{Delilah E.~A. Gates}
\email{dgates@g.harvard.edu}
\affiliation{Center for the Fundamental Laws of Nature, Harvard University, Cambridge, MA 02138, USA}
\author{Shahar Hadar}
\email{shaharhadar@g.harvard.edu}
\affiliation{Center for the Fundamental Laws of Nature, Harvard University, Cambridge, MA 02138, USA}
\author{Alexandru Lupsasca}
\email{lupsasca@princeton.edu}
\affiliation{Princeton Gravity Initiative, Princeton University, Princeton, NJ 08544, USA}

\begin{abstract}
We consider monochromatic and isotropic photon emission from circular equatorial Kerr orbiters.  We derive analytic expressions for the photon escape probability and the redshift-dependent total flux collected on the celestial sphere as a function of emission radius and black hole parameters.  These calculations crucially involve the critical curve delineating the region of photon escape from that of photon capture in each emitter's sky.  This curve generalizes to finite orbital radius the usual Kerr critical curve and displays interesting features in the limit of high spin, which we investigate by developing a perturbative expansion about extremality.  Although the innermost stable circular orbit appears to approach the event horizon for very rapidly spinning black holes, we find in this regime that the photon escape probability tends to $5/12+1/(\sqrt{5}\pi)\arctan\sqrt{5/3}\approx54.65\%$.  We also obtain a simple formula for the flux distribution received on the celestial sphere, which is nonzero.  This confirms that the near-horizon geometry of a high-spin black hole is in principle observable.  These results require us to introduce a novel type of near-horizon double-scaling limit.  We explain the dip observed in the total flux at infinity as an imprint of the black hole: the black hole ``bite''.
\end{abstract}

\maketitle

\section{Introduction}

\begin{figure*}
\begin{tabular}{cc}
	\includegraphics[width=\columnwidth]{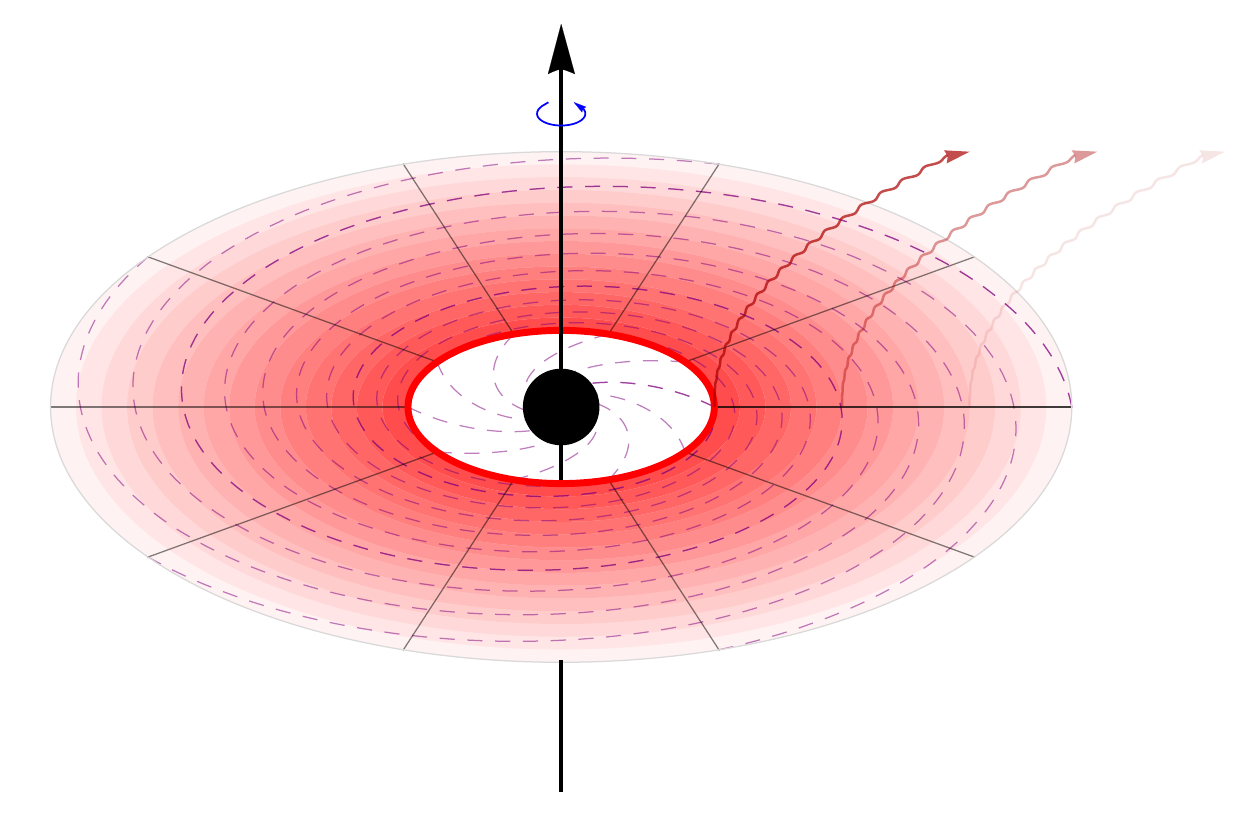} & \includegraphics[width=\columnwidth]{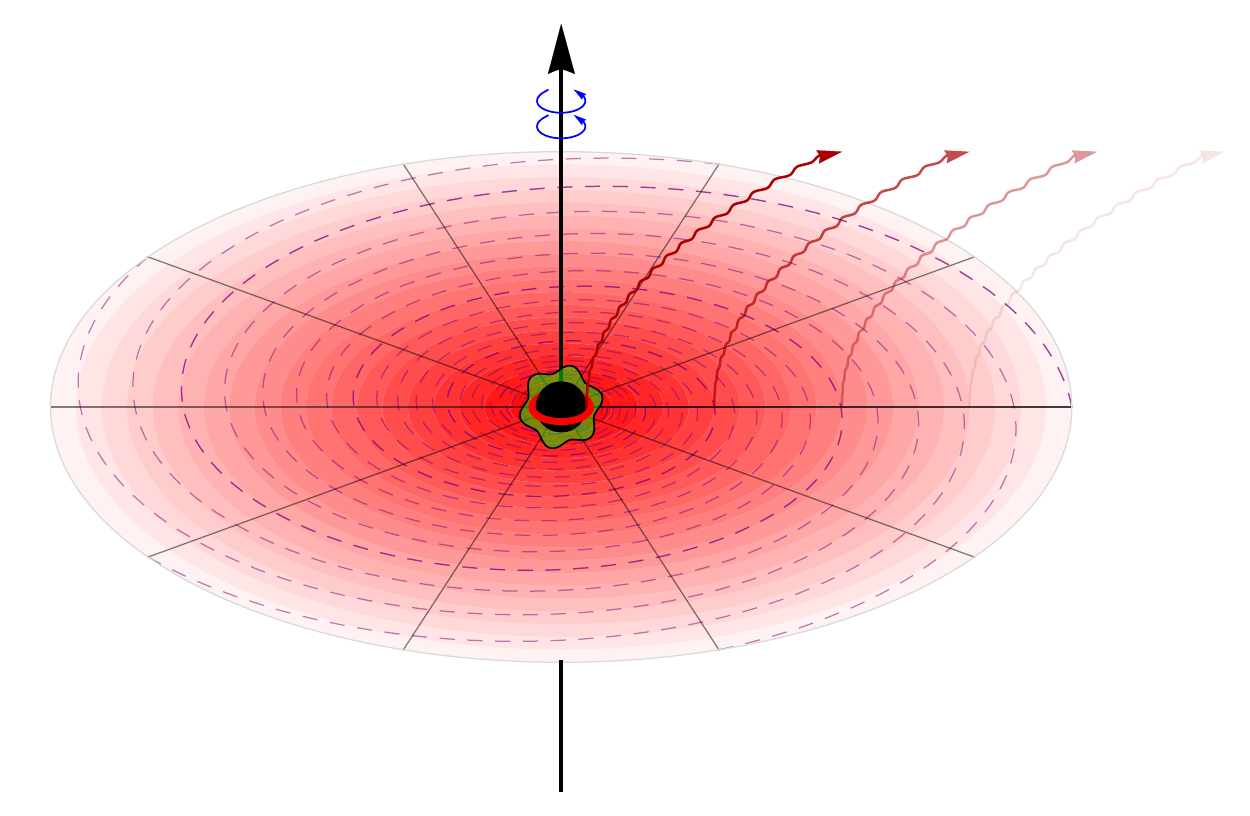} \\
	\includegraphics[width=\columnwidth]{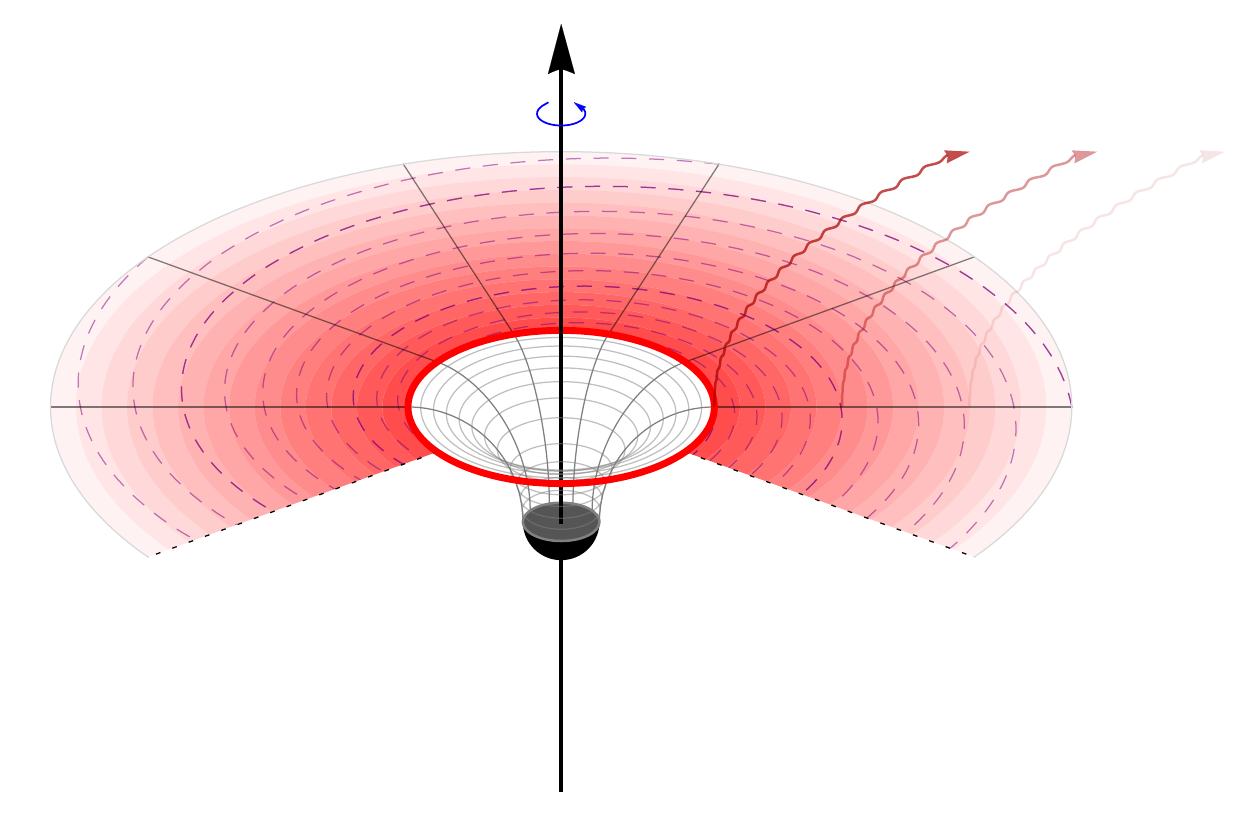} & \includegraphics[width=\columnwidth]{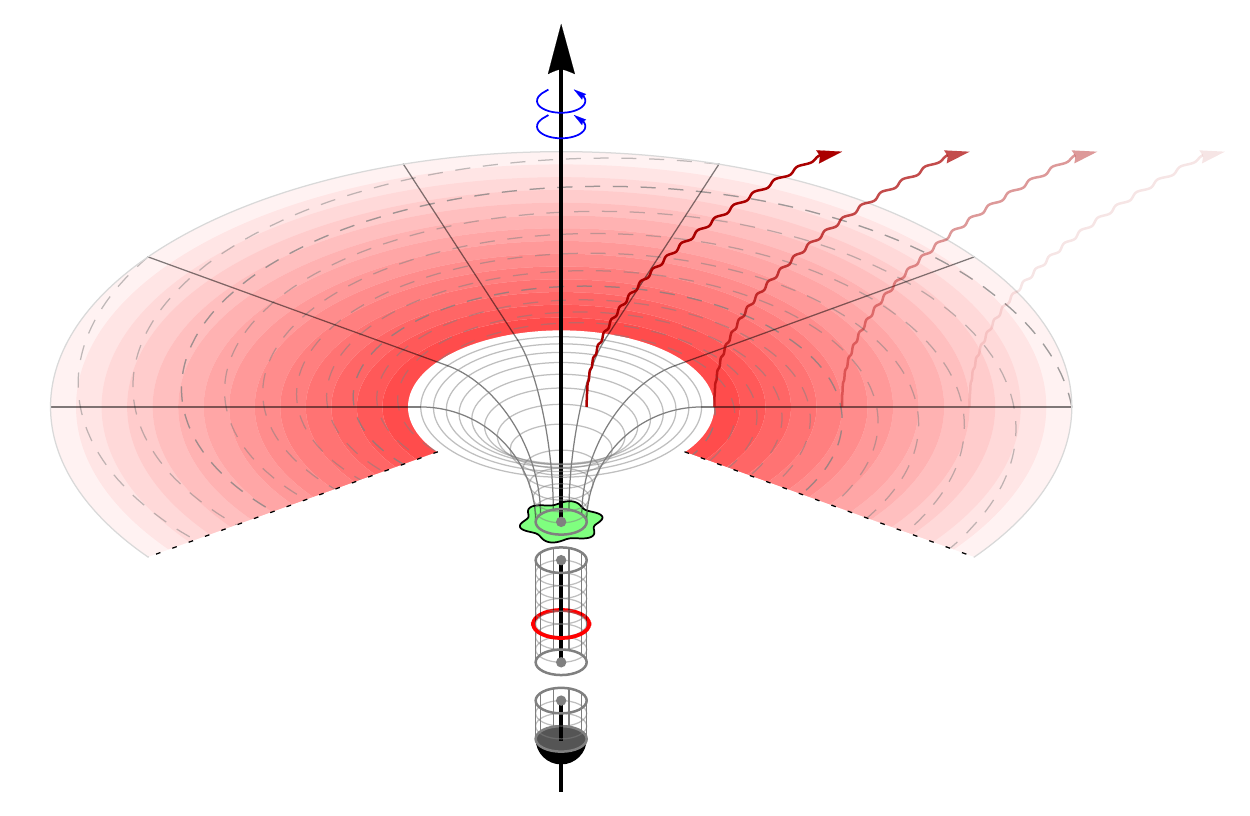}
\end{tabular}
\caption{Top: equatorial disk accreting onto a black hole of generic spin (left) and very high spin (right).  Stable orbits extend down to the ISCO radius [Eq.~\eqref{eq:ISCO}], where the disk terminates.  Bottom: as the BH spin approaches extremality ($a\to M$), the region of spacetime in the vicinity of its event horizon develops an increasingly deep throat.  Its length diverges logarithmically in the deviation from extremality $\kappa=\sqrt{1-a^2/M^2}\ll1$, and the resulting NHEK geometry (App.~\ref{app:NHEK}) exhibits an emergent conformal symmetry enhancing time-translations to include dilations.  The ISCO, and therefore the innermost part of the disk, penetrate deep in the throat as $\kappa\to0$.  Emission from NHEK orbiters displays critical behavior: the photon escape probability [Eq.~\eqref{eq:EscapeProbabilityNHEK}] reaches the same nonvanishing fixed point for all NHEK radii, as does the flux received at infinity [Eq.~\eqref{eq:CriticalFlux}].}
\label{fig:NHEK}
\end{figure*}

Last year, the Event Horizon Telescope collaboration released the first horizon-scale image of a black hole (BH) \cite{EHT2019a}, a remarkable achievement that opens the door to future probes of strong gravity through electromagnetic observations.  It is of great interest, therefore, to develop an intuitive understanding of the characteristic signatures of radiation originating from the vicinity of a BH.  Precise features of the near-horizon emission generically depend on the astrophysical details of the source, requiring numerical simulations to obtain specific predictions.\footnote{A notable exception is the photon ring \cite{Luminet1979,Falcke2000,Beckwith2005} and its universal (i.e., matter-independent) subring structure \cite{Gralla2019,Johnson2019,GrallaLupsasca2020a,Himwich2020,GrallaLupsasca2020d,Hadar2020}.}  In this paper, we study the propagation of radiation from a thin disk composed of isotropic emitters on circular equatorial orbits around a Kerr black hole (Fig.~\ref{fig:NHEK} top).  This simple toy model has the benefit of being analytically tractable, and we hope some of the properties we derive will provide useful heuristics that hold more generally.

While our treatment applies to arbitrary BH spin and orbital radius for the emitters, we give special consideration to emission emanating from the region of spacetime just outside of a near-extremal BH.  Such rapidly rotating BHs exhibit a striking feature: their near-horizon region is stretched into a throat-like geometry described by the Near-Horizon Extremal Kerr (NHEK) metric, which forms a nondegenerate solution to the Einstein equations in its own right \cite{Bardeen1999}.  The NHEK region of spacetime enjoys an emergent conformal symmetry \cite{Bardeen1999,Guica2009} that governs near-horizon physics and has been successfully employed to analytically study a variety of astrophysical processes (see, e.g., Refs.~\cite{Hadar2014,Lupsasca2014,Gralla2016,Compere2017,Gates2018,Kapec2020} and references therein).

Importantly, in the high-spin regime, the innermost stable circular orbit (ISCO) sits deep within the NHEK.  This fact allows a significant portion of the accretion disk to permeate the throat region and probe the extreme conditions therein (Fig.~\ref{fig:NHEK} bottom), with interesting and potentially observable consequences \cite{Gralla2017,Lupsasca2018} that we revisit.

A question we study in detail is: what is the probability for a photon emitted from our radiating orbiters to reach asymptotic null infinity?  As the orbital radius is taken to infinity, one would expect all the emission to escape and this probability to tend to unity.  On the other hand, as the orbiter approaches the horizon, one may naively expect all the emission to be captured by the BH.  For plunging trajectories, both expectations are correct: most of the radiation from radially infalling emitters is beamed into the BH, creating a brightness deficit known as the shadow \cite{Falcke2000,Narayan2019}.  For orbiting emitters, however, we will see that the situation is more subtle.  In order to compute the escape probability, we consider the curve in each emitter's sky that separates directions leading to capture from those leading to escape.  This so-called Kerr critical curve was first computed for distant observers by Bardeen \cite{Bardeen1973}; here, we generalize it to observers at finite radius in the geometry.  Physically, this curve corresponds to light rays that asymptote to bound photon orbits, and it can be viewed as the apparent shape of the BH from an orbiter's perspective.  What does it look like for observers near the BH, including those that are deep in the NHEK and maximally co-rotating with the BH?

Our paper begins with a short review of some essential details of the Kerr geometry and its geodesics (Sec.~\ref{sec:KerrGeodesics}).  Next, we define the emitter sky and introduce angles to parametrize it (Fig.~\ref{fig:Angles}).  We then proceed to compute the critical curve on this sphere (Sec.~\ref{sec:SkyCriticalCurve}), and its area-preserving projection to the plane (Fig.~\ref{fig:BacksideProjection}), which allows us to derive the escape probability (Sec.~\ref{sec:EscapeProbability}).  In the NHEK regime, this procedure is only made possible by introducing a novel type of near-horizon limit, involving a double scaling of both emitter and photon orbit radii.

For near-horizon emitters at high spin, we find that our naive expectation fails: remarkably, the escape probability reaches a critical fixed point of $\sim54.65\%$ (Fig.~\ref{fig:EscapeProbability}).  This was first numerically observed last year \cite{Igata2020}.  Here, we analytically determine the critical probability to be
\begin{align}
    \label{eq:CriticalProbability}
    \mathcal{P}_e=\frac{5}{12}+\frac{\arctan\sqrt{5/3}}{\sqrt{5}\pi},
\end{align}
independently of the NHEK radius.  Thus, every NHEK orbiter sees the BH as having the same specific apparent shape (dotted red curve in Fig.~\ref{fig:BacksideProjection}) that fills almost, but not quite, half of its local sky.  This scale-invariance is a reflection of the dilation symmetry of the throat.  The fact that the probability \eqref{eq:CriticalProbability} is nonzero also implies that the throat geometry is potentially visible.

To explore this possibility, we then ask (Sec.~\ref{sec:FluxCurves}): what is the total spectral flux (flux distribution as a function of observation frequency) received at infinity (collected on the entire celestial sphere) from emission at a given radius?  We answer this question by defining a different projection of the local emitter sky that is weighted by the observed redshift at infinity (Fig.~\ref{fig:RedshiftProjection}).  Integrating over the escape region yields the total flux received at infinity, shown as a function of redshift in Fig.~\ref{fig:FluxCurves} for a representative sample of orbital radii.  A prominent feature of these spectral flux curves immediately stands out: they all display a small dip that spoils an otherwise perfectly linear profile.  Our calculation reveals this deficit to be a signature of photon capture by the BH, which takes a ``bite'' out of the curve.  Similar bites also appear in standard flux curves measured at a specific observer inclination (see, e.g., Figs.~1 of Refs.~\cite{Brenneman2006,Gates2020}).  Here, the bite is precisely the Kerr critical curve, seen not in the sky but in the radiated energy spectrum!

In the NHEK regime, we encounter simplifications that allow us to derive a short analytic formula for the flux [Eq.~\eqref{eq:CriticalFlux} below], which we plot in Fig.~\ref{fig:CriticalFluxCurve} (blue curve).  Again, this result is independent of NHEK radius, providing yet another example of critical behavior.  Moreover, it shows that each NHEK orbiter emits finite flux to infinity, confirming the results obtained in Ref.~\cite{Lupsasca2018} using a different method.\footnote{These results are in apparent contradiction with Fig.~6 of Ref.~\cite{Cunningham1973}, likely because its authors did not account for NHEK contributions.}  Since the depth of the throat diverges as the BH spin approaches extremality, the flux contribution from the NHEK portion of a disk may thus become dominant in the limit of maximal spin (at least in principle, though likely not in practice for realistic spins).

We then turn to finer details of the emission (Sec.~\ref{sec:Unfolding}), studying the angular distribution of emitted photons at the endpoint of their trajectory---either on the celestial sphere or the event horizon of the BH.  The behavior of light rays leaving the emitter reveals an intricate structure of BH lensing \cite{Oliver2015}, in which both the celestial sphere and horizon are unfolded (formally) infinitely many times on the orbiter sky (Fig.~\ref{fig:Unfolding}).\footnote{These plots also show that photons can reach infinity along trajectories that encounter one radial and zero polar turning points.}  We discuss how to derive the angular flux distribution on the celestial sphere from the emitter frame, and perform a detailed comparison (Fig.~\ref{fig:DiskEnvelope}) to previous work \cite{Lupsasca2018} that studied NHEK emissions in the observer frame (orange curve in Fig.~\ref{fig:CriticalFluxCurve}).

We conclude with a discussion of the high-spin perturbative expansion and its extension to subleading order beyond NHEK.  We argue that, in order to approximate the critical curve for near-horizon emitters to higher and higher orders, one needs to consider more and more subregions of the throat geometry (NHEK ``bands'') with different scaling behaviors.  Each NHEK band resolves a different segment of the critical curve associated with photon orbits at different depths in the throat, with the full curve obtained by gluing together these segments at intermediate points along it.\footnote{In App.~\ref{app:NHEK}, we review the NHEK geometry and check that it cannot by itself compute the emitter sky critical curve, which always requires matching across multiple regions (NHEK bands).}  Such a subleading approximation to the critical curve is shown in Fig.~\ref{fig:SubleadingCriticalCurve}.  This suggests the existence of a map between the critical curve and different near-horizon scaling dimensions, whose study we defer to future work.

\section{Kerr geodesics}
\label{sec:KerrGeodesics}

In this section, we describe some features of the Kerr spacetime that will be needed for our computations.  We begin with a brief general discussion of geodesic motion in the Kerr geometry, and then focus on two special classes of geodesics: timelike circular equatorial orbits, which will describe the motion of our emitters, and (unstable) spherical photon orbits, which make up the photon shell.

\subsection{Kerr geometry and geodesics}

Astrophysical BHs are described by the Kerr metric, which depends on only two BH parameters: the mass $M$ and angular momentum $J=aM$.  In Boyer-Lindquist coordinates $(t,r,\theta,\phi)$, the Kerr line element is
\begin{subequations}
\label{eq:Kerr}
\begin{gather}
	ds^2=-\frac{\Delta}{\Sigma}\pa{\ed t-a\sin^2{\theta}\ed\phi}^2+\frac{\Sigma}{\Delta}\ed r^2+\Sigma\ed\theta^2\notag\\
	+\frac{\sin^2{\theta}}{\Sigma}\br{\pa{r^2+a^2}\ed\phi-a\ed t}^2,\\
	\Delta=r^2-2Mr+a^2,\quad
	\Sigma=r^2+a^2\cos^2{\theta}.
\end{gather}
\end{subequations}
In the Kerr spacetime, a particle of mass $\mu$ with affinely parametrized trajectory $x^\nu(\tau)$ has four-momentum
\begin{align}
	\label{eq:KerrMomentum}
	p_\nu\ed x^\nu=-E\ed t\pm_r\frac{\sqrt{\mathcal{R}(r)}}{\Delta(r)}\ed r\pm_\theta\sqrt{\Theta(\theta)}\ed\theta+L\ed\phi
\end{align}
given in terms of the radial and angular potentials
\begin{align}
    \label{eq:RadialPotential}
	\mathcal{R}(r)&=\br{E\pa{r^2+a^2}-aL}^2\notag\\
	&\quad-\Delta(r)\br{Q+\pa{L-aE}^2+\mu^2r^2},\\
	\label{eq:AngularPotential}
	\Theta(\theta)&=Q+a^2\pa{E^2-\mu^2}\cos^2{\theta}-L^2\cot^2{\theta},
\end{align}
where the quantities $(E,L,Q)$ respectively denote the energy, angular momentum parallel to the axis of symmetry, and Carter constant of the particle, all of which are conserved along a geodesic.  The particle's trajectory is determined by its initial position $x^\nu(0)$, together with its conserved quantities $(\mu,E,L,Q)$ and the signs $\pm_r$ and $\pm_\theta$ denoting its initial polar and radial directions of motion.  The existence of these four conserved quantities ensures that Kerr geodesic motion is completely integrable \cite{Carter1968}.

For null geodesics ($\mu=0$), it is convenient to define the energy-rescaled angular momentum and Carter constant,
\begin{align}
    \label{eq:EnergyRescaledQuantities}
	\lambda=\frac{L}{E},\quad
	\eta=\frac{Q}{E^2},
\end{align}
as this allows for the energy $E$ to completely factor out of the four-momentum \eqref{eq:KerrMomentum}.  As a result, the trajectory of a massless particle is independent of its energy and characterized only by its energy-rescaled conserved quantities $(\lambda,\eta)$.  The null geodesic equation can be solved analytically using elliptic integrals \cite{Rauch1994,Vazquez2004,Dexter2009,Hackmann2010}, and a convenient form of the solution was explicitly derived in Ref.~\cite{GrallaLupsasca2020b}.

\subsection{Circular equatorial orbiters}
\label{sec:CircularOrbiters}

The Kerr geometry admits timelike circular equatorial geodesics at any orbital radius $r=r_s$, with conserved quantities obtained by solving $\Theta(\pi/2)=\Theta'(\pi/2)=0$ and $\mathcal{R}(r_s)=\mathcal{R}'(r_s)=0$.  The first condition requires $Q=0$, while the latter implies \cite{Bardeen1972}
\begin{subequations}
\begin{align}
	E&=\frac{\mu}{\xi_s}\pa{r_s^{3/2}-2M\sqrt{r_s}\pm a\sqrt{M}},\\
	L&=\pm\frac{\mu\sqrt{M}}{\xi_s}\pa{r_s^2\mp2a\sqrt{Mr_s}+a^2},\\
	\xi_s&=\sqrt{r_s^3-3Mr_s^2\pm2a\sqrt{M}r_s^{3/2}},
\end{align}
\end{subequations}
with the upper and lower sign corresponding to prograde and retrograde orbits, respectively.  These orbits have an angular velocity $\Omega_s$ and four-velocity $u_s^\nu$ given by
\begin{align}
	\Omega_s&=\pm\frac{\sqrt{M}}{r_s^{3/2}\pm a\sqrt{M}},\\
	u_s&=u_s^t\pa{\pd_t+\Omega_s\pd_\phi},\quad
	u_s^t=\frac{r_s^{3/2}\pm a\sqrt{M}}{\xi_s}.
\end{align}
The orbital motion is strictly stable as long as $\mathcal{R}''(r_s)<0$ and becomes marginally stable at the ISCO radius
\begin{subequations}
\label{eq:ISCO}
\begin{align}
	r_\mathrm{ms}^\pm&=M\br{3+Z_2\mp\sqrt{\pa{3-Z_1}\pa{3+Z_1+2Z_2}}},\\
	Z_1&=1+\sqrt[3]{1-a_\star^2}\pa{\sqrt[3]{1+a_\star}+\sqrt[3]{1-a_\star}},\\
	Z_2&=\sqrt{3a_\star^2+Z_1^2},\quad
	a_\star=\frac{a}{M},
\end{align}
\end{subequations}
such that $\mathcal{R}''(r_\mathrm{ms}^\pm)=0$.  The rotation of the BH induces ``frame-dragging'' at a characteristic angular velocity
\begin{align}
	\omega=-\frac{g_{t\phi}}{g_{\phi\phi}}
	=\frac{2aMr}{\pa{r^2+a^2}^2-a^2\Delta(r)\sin^2{\theta}}.
\end{align}
Physics in a local frame is described using a tetrad with the observer's four-velocity as its time leg.  For a circular orbiter, it is convenient to define an orthonormal tetrad with components aligned with the $(r,\theta)$ directions,
\begin{subequations}
\label{eq:OrthornomalFrame}
\begin{align}
	\mathbf{e}_{(t)}&=u_s,\\
	\mathbf{e}_{(r)}&=\sqrt{1-\frac{2M}{r_s}+\frac{a^2}{r_s^2}}\pd_r,\\
	\mathbf{e}_{(\theta)}&=\frac{1}{r_s}\pd_\theta,\\
	\mathbf{e}_{(\phi)}&=v_su_s^t\pa{\pd_t+\omega_s\pd_\phi}+\sqrt{\frac{\omega_sr_s}{2aM\pa{1-v_s^2}}}\pd_\phi,
\end{align}
\end{subequations}
where $\omega_s\equiv\omega(r=r_s,\theta=\pi/2)$, and
\begin{align}
	v_s=\frac{\pm\sqrt{M}\big(r_s^2\mp2a\sqrt{Mr_s}+a^2\big)}{\sqrt{\Delta(r_s)}\big(r_s^{3/2}\pm a\sqrt{M}\big)}.
\end{align}
This quantity is the orbital velocity in the $\phi$ direction of a circular orbiter, as measured by a ``locally nonrotating'' observer tracing an equatorial worldline $\phi=\omega_st$ at the same radius \cite{Bardeen1972}.  The orthonormal frame \eqref{eq:OrthornomalFrame} obeys
\begin{align}
    \label{eq:TetradConditions}
	g_{\mu\nu}\mathbf{e}_{(a)}^\mu\mathbf{e}_{(b)}^\nu=\eta_{(a)(b)},\quad
	\eta^{(a)(b)}\mathbf{e}_{(a)}^\mu\mathbf{e}_{(b)}^\nu=g^{\mu\nu},
\end{align}
where $\eta^{(a)(b)}=\mathrm{diag}\pa{-1,1,1,1}$, everywhere along its orbit.  Frame components of four-vectors $V^\mu$ are given by
\begin{align}
	V^{(a)}=\eta^{(a)(b)}\mathbf{e}_{(b)}^\mu V_\mu.
\end{align}

\subsection{Bound photon orbits}

The Kerr geometry also admits spherical bound photon orbits of fixed radius $r=\rt$, with conserved quantities obtained by solving $\mathcal{R}(\rt)=\mathcal{R}'(\rt)=0$ \cite{Bardeen1973,Teo2003}.  These conditions can only be satisfied for orbital radii in the range $\rt_-\le\rt\le\rt_+$, where
\begin{align}
	\rt_\pm=2M\br{1+\cos\pa{\frac{2}{3}\arccos\pa{\pm\frac{a}{M}}}},
\end{align}
and they require the conserved quantities $(\lambda,\eta)$ to be tuned to their critical values
\begin{subequations}
\label{eq:CriticalParameters}
\begin{align}
	\tilde{\lambda}(\rt)&=a+\frac{\rt}{a}\br{\rt-\frac{2 \Delta(\rt)}{\rt-M}},\\
	\tilde{\eta}(\rt)&=\frac{\rt^{3}}{a^2}\br{\frac{4M\Delta(\rt)}{\pa{\rt-M}^2}-\rt}.
\end{align}
\end{subequations}
The bound photon orbits are always unstable because $\mathcal{R}''(\rt)>0$.  They span a region of spacetime known as the photon shell and are responsible for the photon ring present in BH images \cite{Gralla2019,Johnson2019,GrallaLupsasca2020a,Himwich2020,GrallaLupsasca2020d}.  The critical parameters \eqref{eq:CriticalParameters} define a critical locus \cite{Chandrasekhar1983,GrallaLupsasca2020c}
\begin{align}
    \label{eq:CriticalLocus}
	\mathcal{C}=\cu{\pa{\tilde{\lambda}(\rt),\tilde{\eta}(\rt)}\Big|\rt_-\le\tilde r\le\rt_+},
\end{align}
in $(\lambda,\eta)$-space (see, e.g., Fig.~3 of Ref.~\cite{GrallaLupsasca2020b}).  In the remainder of this paper, if a quantity $X(r,\lambda,\eta)$ is adorned with a tilde, then it is to be evaluated on a critical radius and its associated critical parameters; that is,
\begin{align}
    \tilde{X}=X\pa{\rt,\tilde{\lambda}(\rt),\tilde{\eta}(\rt)}.
\end{align}

\section{Critical curve in orbiter sky}
\label{sec:SkyCriticalCurve}

In this section, we study the sky of timelike emitters on stable ($r_s\ge r_\mathrm{ms}^\pm$) circular equatorial orbits (Sec.~\ref{sec:CircularOrbiters}).  The sky is naturally parametrized by light rays with different conserved quantities $(\lambda,\eta)$, which we relate to various local angles of emission.  This allows us to map the critical locus \eqref{eq:CriticalLocus} into a closed curve in the orbiter sky delineating the region of photon capture, in which emitted photons eventually cross the horizon, from that of photon escape, in which they eventually reach asymptotic null infinity.  This generalizes to finite-distance observers the Kerr critical curve first derived for far observers by Bardeen \cite{Bardeen1973}.  We take special care to examine the extremal limit $a\to M$, which presents special difficulties, particularly when the orbiter approaches the prograde ISCO radius $r_\mathrm{ms}^+$.

\subsection{Orbiter sky}

\begin{figure}
    \centering
	\includegraphics[width=\columnwidth]{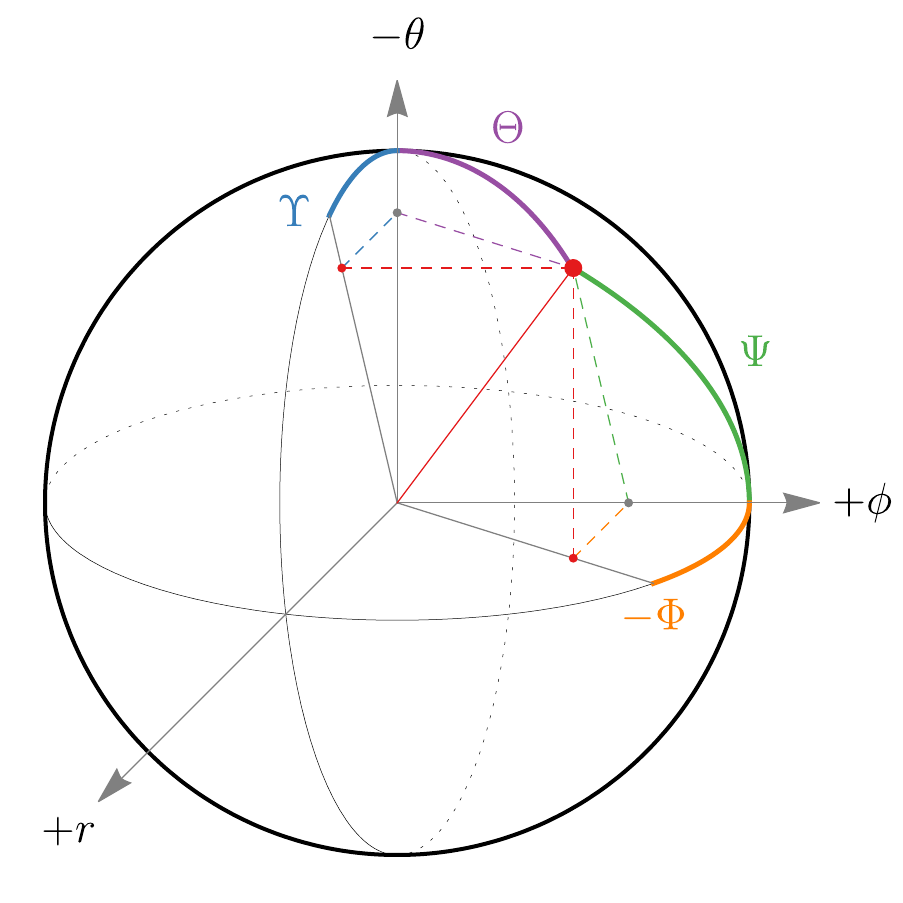}
	\caption{Local angles parametrizing the orbiter sky.}
	\label{fig:Angles}
\end{figure}

The sky of a circular equatorial orbiter at radius $r_s$ may be parametrized using the angles depicted in Fig.~\ref{fig:Angles}.  Following Cunningham and Bardeen \cite{Cunningham1973}, let $\Theta\in[0,\pi]$ denote the polar angle from the $-\theta$ direction (the local zenith) and $\Psi\in[0,\pi]$ the polar angle from the $+\phi$ direction, which is either the direction of motion or its opposite, according to whether the orbit is prograde or retrograde, respectively.\footnote{These angles were also used in Ref.~\cite{Gralla2017}, where $\Psi$ was denoted $\Phi$.}  A photon with conserved quantities $(\lambda,\eta)$ is emitted in the direction
\begin{align}
    \label{eq:PolarAngle}
	\cos{\Theta}&=-\frac{p^{(\theta)}}{p^{(t)}}
	=\mp_\theta\frac{\xi_s\sqrt{\eta}}{r_s\br{r_s^{3/2}\pm\sqrt{M}\pa{a-\lambda}}},\\
    \label{eq:ForwardCosine}
	\cos{\Psi}&=\frac{p^{(\phi)}}{p^{(t)}}
	=\frac{\frac{\omega_sr_s}{2aM}\sqrt{\Delta(r_s)}\lambda-v_s\pa{1-\omega_s\lambda}}{1-\Omega_s\lambda}.
\end{align}
The angles $(\Theta,\Psi)$ are not good coordinates on the sphere because they are both polar: an allowed joint choice of $(\Theta,\Psi)$ generically defines two light rays with the same $(\lambda,\eta)$ but opposite signs $\pm_r$.  Viewing $\Theta$ as the usual polar angle, the standard azimuthal angle is $\Phi\in(-\pi,\pi]$, the projection of $\Psi$ into the equatorial plane, which obeys
\begin{align}
    \cos{\Phi}=\frac{\cos{\Psi}}{\sin{\Theta}},
\end{align}
as can be inferred from Fig.~\ref{fig:Angles}.  More explicitly,
\begin{align}
    \Phi=\mp_r\arccos\pa{\frac{\cos{\Psi}}{\sqrt{1-\cos^2{\Theta}}}}.
\end{align}
The angles $(\Theta,\Phi)$ are standard coordinates on the sphere with area element $\ed\Omega=\sin{\Theta}\ed\Theta\wedge\ed\Phi$, but we will find it more convenient to use the $+\phi$ direction as the local zenith.  That is, we use $\Psi$ as a polar angle and use the projection of $\Theta$ into the plane dividing the forward and backward hemispheres as an azimuthal angle $\Upsilon\in(-\pi,\pi]$,
\begin{align}
    \label{eq:Upsilon}
	\Upsilon=\pm_r\arccos\pa{\frac{\cos{\Theta}}{\sqrt{1-\cos^2{\Psi}}}}.
\end{align}
This last relation was inferred by using Fig.~\ref{fig:Angles} to see that
\begin{align}
    \cos{\Upsilon}=\frac{\cos{\Theta}}{\sin{\Psi}}.
\end{align}
The angles $(\Psi,\Upsilon)$ also define standard coordinates on the sphere with area element $\ed\Omega=\sin{\Psi}\ed\Psi\wedge\ed\Upsilon$.

Finally, note that the angular potential \eqref{eq:AngularPotential} takes the form $E^2\pa{\eta+a^2\cos^2{\theta}-\lambda^2\cot^2{\theta}}$ for null geodesics.  Since it must be nonnegative everywhere along a physical trajectory, only light rays with $\lambda=0$ can reach the BH spin axis, where $\sin{\theta}=0$.  Moreover, light rays emitted from the equator with $\eta=0$ have vanishing angular potential and therefore $p^\theta=0$, so they remain in the equatorial plane forever.  Hence, it is natural to define a ``direction to the black hole center'' by the light ray with $\lambda=\eta=0$ and $\pm_r=-1$; that is, with emission angles
\begin{align}
    \label{eq:BlackHoleCenter}
    \Theta_\bullet=-\Upsilon_\bullet
    =\frac{\pi}{2},\quad
    \Psi_\bullet=\Phi_\bullet
    =\arccos\pa{-v_s}.
\end{align}

\subsection{Critical curve for generic spin and orbital radius}

The critical curve in the orbiter sky is the closed curve
\begin{align}
    \label{eq:SkyCriticalCurve}
	\mathcal{C}=\cu{\pa{\tilde{\Psi}(\rt),\tilde{\Upsilon}(\rt)}\Big|\rt_-\le\tilde r\le\rt_+},
\end{align}
where $\pm_r=\sign\pa{\rt-r_s}$ along the critical curve, so
\begin{align}
    \tilde{\Upsilon}=\sign\pa{\rt-r_s}\arccos\pa{\frac{\cos{\tilde{\Theta}}}{\sqrt{1-\cos^2{\tilde{\Psi}}}}}.
\end{align}
The critical curve divides the sky into two regions: one in which light rays connect to the horizon, and another in which they connect to asymptotic null infinity.  That is, it delineates the region of photon capture from that of photon escape.  Photons emitted in a direction precisely on the critical curve asymptote to bound photon orbits in the photon shell.  Note that the critical curve is a theoretical curve and not directly observable in itself.  Nonetheless, it is expected to be significant in many astrophysical scenarios, as strongly lensed light rays must appear exponentially close to it, and this may result in a brightness enhancement: the photon ring \cite{Johnson2019,GrallaLupsasca2020a,GrallaLupsasca2020d}.

Escaping photons reach infinity with redshift equal to the ratio of observed energy to local emission energy,
\begin{align}
    \label{eq:Redshift}
	g&=\frac{E}{p^{(t)}}
	=\frac{\xi_s}{r_s^{3/2}\pm\sqrt{M}\pa{a-\lambda}}\\
	\label{eq:RedshiftPsi}
	&=\frac{r_s^{3/2}-2M\sqrt{r_s}\pm\sqrt{M}\pa{a+\sqrt{\Delta(r_s)}\cos{\Psi}}}{\xi_s},
\end{align}
where in the last line we used Eq.~\eqref{eq:ForwardCosine} to relate the observed redshift at infinity to the emission angle $\Psi$ in the rest frame of the source.  Since $g(\Psi)$ is monotonic in $\Psi$,
\begin{align}
    \label{eq:RedshiftMonotonicity}
    \pd_\Psi g(\Psi)=\mp\frac{\sqrt{M\Delta(r_s)}}{\xi_s}\sin{\Psi}\lessgtr0,
\end{align}
the maximum attainable blueshift is obtained by photons emitted in the direction of orbital motion, namely, $\Psi=0$ (the $+\phi$ direction) for prograde orbiters or $\Psi=\pi$ (the $-\phi$ direction) for retrograde orbiters.  Since such photons are emitted into the equatorial plane, they have $\cos{\Theta}=0$ and therefore $\eta=0$, as Eqs.~\eqref{eq:PolarAngle} and \eqref{eq:Redshift} imply that
\begin{align}
    \cos{\Theta}=\mp_\theta\frac{g\sqrt{\eta}}{r_s}.
\end{align}
It then follows from Eq.~\eqref{eq:KerrMomentum} that these photons must always remain in the equatorial plane.  We will graphically show (Fig.~\ref{fig:BacksideProjection} below) that they always escape to asymptotic infinity.  Hence, maximally blueshifted photons always reach the celestial sphere at the equator, producing interesting and potentially observable effects \cite{Gates2020}.

Note that $g(\Psi_0)=0$ for
\begin{align}
    \cos{\Psi_0}=\mp\frac{r_s^{3/2}-2M\sqrt{r_s}\pm a\sqrt{M}}{\sqrt{M\Delta(r_s)}},
\end{align}
so that the redshift factor becomes (formally) negative for $\Psi\gtrless\Psi_0$, according to whether the orbiter is prograde or retrograde, respectively [Eq.~\eqref{eq:RedshiftMonotonicity}].  Negative redshift requires the local emitted energy $p^{(t)}$ to be negative, which can only happen inside the Kerr ergosphere, where the Killing vector $\pd_t$ becomes spacelike.  Mathematically, this follows from the fact that the angle $\Psi_0$ is only physical if $\ab{\cos{\Psi_0}}\le1$, which happens only for $r_s\le2M$, that is, once the emitter enters the ergosphere, whose equatorial radius is always at $r=2M$ regardless of BH spin.  Stable retrograde orbiters are always outside the ergosphere, while the prograde ISCO enters it when $r_\mathrm{ms}^+=2M$, or
\begin{align}
    \frac{a}{M}=\frac{2\sqrt{2}}{3}\approx94.3\%.
\end{align}
Thus, it is only for prograde orbiters and spins above this bound that part of the backward hemisphere has a (formally) negative redshift \eqref{eq:RedshiftPsi}.  Since observed energy at infinity must be positive, light with negative redshift can never possibly escape and is therefore captured.  On the other hand, light with positive redshift does not necessarily escape.  It only does so if it is emitted in the escape region of the orbiter sky.  The largest angle of emission for which photons could possibly escape is thus given by
\begin{align}
    \label{eq:PsiMax}
    \Psi_\mathrm{max}=
    \begin{cases}    
        \Psi_0&\text{ if $r_s<2M$ (prograde only)},\\
        \pi&\text{ otherwise}.
    \end{cases}
\end{align}
Hence, the redshift of escaping photons is bounded by $\hat{g}\lessgtr g\lessgtr g_0$, where the upper/lower inequalities correspond to prograde/retrograde orbiters, respectively, and
\begin{align}
   g_0=g(\Psi=0),\quad
   \hat{g}(\Psi=\Psi_\mathrm{max}).
\end{align}
The redshift of a critical photon is given by
\begin{align}
    \label{eq:CriticalRedshift}
    \tilde{g}(\rt)=g\pa{\Psi=\tilde{\Psi}(\rt)}, 
\end{align}
which takes values in the range
\begin{align}
    \label{eq:RedshiftBounds}
    \hat{g}\lessgtr\tilde{g}_+\lessgtr\tilde{g}\lessgtr\tilde{g}_-\lessgtr g_0,\quad \tilde{g}_\pm&=\tilde{g}(\rt_\pm),
\end{align}
with the upper/lower inequalities again corresponding to prograde/retrograde orbiters.  Note that for a prograde orbiter, the smaller radii $\rt$ produce larger values of $\tilde{g}$.

\subsection{Critical curve for infinite orbital radius}

As the orbital radius diverges, the critical curve shrinks to a point $(\Theta_\bullet,\Phi_\bullet)$ in the orbiter sky, since
\begin{align}
    \sin\pa{\Phi-\Phi_\bullet}&=-\frac{\lambda}{r_s}+\O{\frac{1}{r_s^2}},\\
    \sin\pa{\Theta-\Theta_\bullet}&=\mp_\theta\frac{\sqrt{\eta}}{r_s}+\O{\frac{1}{r_s^2}}.
\end{align}
We can blow up a neighborhood of that point as $r_s\to\infty$ and define Cartesian coordinates in the resulting sky as
\begin{align}
    \alpha_s&=\lim_{r_s\to\infty}r_s\sin\pa{\Phi-\Phi_\bullet}
    =-\lambda,\\
    \beta_s&=\lim_{r_s\to\infty}r_s\sin\pa{\Theta-\Theta_\bullet}
    =\mp_\theta\sqrt{\eta}.
\end{align}
If we flip the sign $\pm_\theta=\sign\pa{p_s^\theta}$ of $\beta$ to describe received rather than emitted light rays, then these coordinates match those previously defined by Bardeen \cite{Bardeen1973},\footnote{Bardeen \cite{Bardeen1973} defines $\alpha=-r_op^{[\phi]}/p^{[t]}$ and $\beta=r_op^{[\theta]}/p^{[t]}$, where $p^{[a]}$ is the four-momentum in a locally nonrotating frame at infinity.}
\begin{subequations}
\label{eq:BardeenCoordinates}
\begin{align}
    \alpha_o&=\lim_{r_o\to\infty}-r_o\frac{p^{(\phi)}}{p^{(t)}}
    =-\frac{\lambda}{\sin{\theta_o}},\\
    \beta_o&=\lim_{r_o\to\infty}r_o\frac{p^{(\theta)}}{p^{(t)}}
    =\pm_\theta\sqrt{\Theta(\theta_o)},
\end{align}
\end{subequations}
in the sky of a distant equatorial observer at $\theta_o=\pi/2$.  As such, when the orbital radius is infinite, the orbiter sky critical curve \eqref{eq:SkyCriticalCurve} reproduces the usual observer sky critical curve.  Here, we have generalized this curve to a one-parameter family of curves labeled by finite orbital $r_s$.  As shown in Refs.~\cite{Gralla2017,GrallaLupsasca2020c}, the extremal limit $a\to M$ of the usual $r_s\to\infty$ critical curve is subtle.  We will now see that this limit is even more subtle for finite $r_s$.

\subsection{Critical curve for extremal spin}
\label{sec:ExtremalCriticalCurve}

The extremal limit $a\to M$ of the critical curve \eqref{eq:SkyCriticalCurve} requires special care due to the emergence of the NHEK throat \cite{Bardeen1999} (see Fig.~\ref{fig:NHEK} and App.~\ref{app:NHEK}).  The extreme Kerr metric, obtained by setting $a=M$ into Eq.~\eqref{eq:Kerr}, fails to capture physics in the near-horizon region $r/M-1\ll1$, which is instead resolved by the NHEK geometry \eqref{eq:NHEK} (see, e.g., Refs.~\cite{Gralla2016,Kapec2020} for astrophysically oriented discussions).  As a consequence, simply setting $a=M$ into Eq.~\eqref{eq:SkyCriticalCurve} does not yield the full closed critical curve in the orbiter sky, but only a portion of it.

To understand why, consider a high-spin BH with 
\begin{align}
    a=M\sqrt{1-\kappa^2},\quad
	0<\kappa\ll1.
\end{align}
In the limit $\kappa\to0$, the Boyer-Lindquist radii of the form
\begin{align}
    r=M\pa{1+\kappa^pR},\quad
    0<p\le 1,
\end{align}
all end up in the throat.  More precisely, only radii in the same ``NHEK band'' $p$ (i.e., with the same near-extremal scaling behavior) end up in the same NHEK geometry; that is, there are infinitely many NHEK bands labeled by a real number $0<p\le1$ \cite{Hadar2014,Kapec2020}.  At high spins \cite{Hadar2017},
\begin{align}
    \label{eq:PhotonShellNHEK}
	\rt_-&=M\br{1+\frac{2}{\sqrt{3}}\kappa+\O{\kappa^2}},\\
	\rt_+&=4M\br{1+\O{\kappa^2}},
\end{align}
so the photon shell stretches from the $p=1$ NHEK band---the deepest part of the throat---all the way to the asymptotically flat region covered by the extreme Kerr metric, straddling all the intermediate $p$-bands in between.  Thus, it should come as no surprise that the part of the photon shell with radii $\tilde{r}-M\sim\kappa^0$ (resolved by the extreme Kerr metric ``$p=0$'') only parametrizes part of the full physical critical curve, while missing the portion corresponding to photon shell radii in the throat.

We will now show how to properly take the extremal limit $\kappa\to0$ to recover the full critical curve in the orbiter sky.  For an orbiter at infinite radius, this amounts to the calculation presented in App.~A of Ref.~\cite{Gralla2017}, where it was found that the missing contribution from near-horizon photon shell radii produces a vertical line in the observer sky \eqref{eq:BardeenCoordinates}, now known as the NHEKline.  Here, we generalize this limit to orbiters at finite radius
\begin{align}
    r_s=M\pa{1+\kappa^qR_s},\quad
    0\le q\le\frac{2}{3},
\end{align}
where the restriction on NHEK bands arises from the condition $r_s\ge r_\mathrm{ms}^+$, together with the observation that
\begin{align}
    \label{eq:ExtremeISCO}
    r_\mathrm{ms}^+=M\br{1+2^{1/3}\kappa^{2/3}+\O{\kappa^{4/3}}},
\end{align}
for prograde orbiters.  For an orbiter in the $p=0$ region covered by the extreme Kerr metric, the limit of the critical curve remains unchanged, and is also the same limit needed to resolve the critical locus \eqref{eq:CriticalLocus} \cite{GrallaLupsasca2020c}.

On the other hand, when the orbiter lies in a NHEK band with $0<q<2/3$, its critical curve splits into two halves, one parametrized by photon shell radii in the deepest part of the throat (the $p=1$ band) and the other by larger radii that are still in the throat, in the intermediate band $p=q/2$.  Thus, to recover the entire critical curve, one must resort to a novel type of double scaling limit, which we further explore in Sec.~\ref{sec:PerturbationTheory} below.

\subsubsection{Extreme Kerr orbiter}

When the orbiter is in extreme Kerr, the critical curve \eqref{eq:SkyCriticalCurve} decomposes into two pieces,
\begin{align}
    \label{eq:CriticalCurveExtremeKerr}
	\mathcal{C}=\mathcal{C}_f\cup\mathcal{C}_n
\end{align}
where $\mathcal{C}_f$ and $\mathcal{C}_n$ correspond to critical light rays that asymptote to bound photon orbits in the portion of the photon shell outside and inside the throat, respectively.

We resolve $\mathcal{C}_f$ by simply taking $a\to M$ in $(\tilde{\Psi},\tilde{\Upsilon})$.  To resolve $\mathcal{C}_n$, we use the near-extreme, near-horizon limit
\begin{align}
    \label{eq:NHEKLimit}
    a=M\sqrt{1-\kappa^2},\quad
    \rt=M\pa{1+\kappa\Rt},\quad
    0<\kappa\ll1,
\end{align}
and obtain, to leading order as $\kappa\to0$,
\begin{subequations}
\begin{align}
    \tilde{\Psi}_n&=\arccos\pa{\frac{\sqrt{M}\br{2\sqrt{Mr_s}\mp (r_s+M)}}{r_s^{3/2}\mp M^{3/2}}},\\
    \tilde{\Upsilon}_n&=-\arccos\pa{\mp_\theta\sqrt{\frac{3M^2}{r_s(r_s+2M)}\pa{1-\frac{\Rt_-^2}{\Rt^2}}}},
\end{align}
\end{subequations}
where the NHEK radius $\tilde{R}$ ranges over
\begin{align}
    \Rt\in&\left[\Rt_-,\infty\right),\quad
    \Rt_-=\frac{2}{\sqrt{3}},
\end{align}
in accordance with Eq.~\eqref{eq:PhotonShellNHEK}.  As such, $\tilde{\Upsilon}_n$ is bounded by
\begin{align}
    \Upsilon_n^\pm=-\arccos\pa{\pm\sqrt{\frac{3M^2}{r_s(r_s+2M)}}}.
\end{align}
Thus, the critical curve for an extreme Kerr orbiter is
\begin{subequations}
\begin{align}
	\mathcal{C}_f&=\cu{\pa{\tilde{\Psi}(\rt),\tilde{\Upsilon}(\rt)}\Big|M\le\rt\le4M},\\
	\mathcal{C}^n&=\cu{\pa{\tilde{\Psi}^n,\tilde{\Upsilon}}\Big|\Upsilon_n^-\leq\tilde{\Upsilon}\leq\Upsilon_n^+}.
\end{align}
\end{subequations}
The critical light rays on $\mathcal{C}_f$ and $\mathcal{C}_n$ have redshift 
\begin{align}
    \tilde{g}_f&=\frac{\sqrt{M}\xi_s}{r_s^{3/2}\sqrt{M}\pm\rt\pa{\rt-2M}},\quad
    \tilde{g}_n=\frac{\xi_s}{r_s^{3/2}\mp M^{3/2}},
\end{align}
where $\xi_s$ is to be evaluated at $a=M$.

\subsubsection{NHEK orbiter}

When the (necessarily prograde)\footnote{The retrograde ISCO radius ranges over $r_\mathrm{ms}^-\in[6M,9M]$ for BH spins $a\in[0,M]$, so the retrograde orbits stay out of the throat.} orbiter is in the throat, the critical curve \eqref{eq:SkyCriticalCurve} decomposes into two pieces,
\begin{align}
	\mathcal{C}=\mathcal{C}^+\cup\mathcal{C}^-,
\end{align}
where both $\mathcal{C}^\pm$ correspond to critical light rays that asymptote to bound photon orbits in the portion of the photon shell inside the throat, with $\rt\gtrless r_s$ respectively.

To resolve $\mathcal{C}^\pm$, we use a near-extreme, near-horizon limit with two different scalings for $\tilde{r}$ and $r_s$:
\begin{subequations}
\label{eq:DoubleLimit}
\begin{align}
	a&=M\sqrt{1-\kappa^2},
	&&0<\kappa\ll1,\\
	\rt&=M\pa{1+\kappa^p\Rt},
	&&0<p\le1,\\
	r_s&=M\pa{1+\kappa^qR_s},
	&&0<q\le\frac{2}{3}.
\end{align}
\end{subequations}
This double scaling limit has fascinating properties explored in Sec.~\ref{sec:PerturbationTheory} below.  As will become clearer then, only specific choices of $p$ and $q$ result in a well-defined limit.  In particular, resolving the critical curve requires
\begin{align}
    p\pa{\mathcal{C}^{+}}=\frac{q}{2},\quad p\pa{\mathcal{C}^{-}}=1.
\end{align}
When $p=1$, the double limit \eqref{eq:DoubleLimit} reduces to the usual single limit in Eq.~\eqref{eq:NHEKLimit}, so $\mathcal{C}^{-}$ may be interpreted as the NHEKline portion of the critical curve in the  orbiter sky.  Note also that $r_s$ must always exceed the prograde ISCO, which has $q=2/3$ and $R_s=2^{1/3}$ [Eq.~\eqref{eq:ExtremeISCO}].  Taking the limit $\kappa\to0$ with $p=q/2$ results in the critical curve $\mathcal{C}^+$:
\begin{subequations}
\label{eq:CplusNHEK}
\begin{align}
    \tilde{\Psi}^+&=\arccos\pa{-\frac{\Rt^2}{2\Rt^2+3R_s}},\quad
    \Rt\in\pa{0,\infty},\\
    \tilde{\Upsilon}^+&=\arccos\pa{\frac{\mp_\theta\sqrt{3}R_s}{\sqrt{\pa{\Rt^2+R_s}\pa{\Rt^2+3R_s}}}}.
\end{align}
\end{subequations}
The same limit but with $p=1$ yields $\mathcal{C}^-$:
\begin{subequations}
\label{eq:CminusNHEK}
\begin{align}
    \tilde{\Psi}^-&=\frac{\pi}{2},\quad
    \Rt\in\left[\tilde{R}_-,\infty\right),\\
    \tilde{\Upsilon}^-&=-\arccos\pa{\pm_\theta\sqrt{1-\frac{\Rt_-^2}{\Rt}}}.
\end{align}
\end{subequations}
Inverting the relations \eqref{eq:CplusNHEK} results in
\begin{subequations}
\begin{align}
    \frac{\tilde{R}^2}{R_s}&=-\frac{3R_s\cos{\tilde{\Psi}^+}}{1+2\cos{\tilde{\Psi}^+}},\\
    \frac{\tilde{R}^2}{R_s}&=\pm\sqrt{1+\frac{3}{\cos^2{\tilde{\Upsilon}^+}}}-2,
\end{align}
\end{subequations}
where only the $+$ root is physical.  Equating these two expressions results in a direct parametrization for $\mathcal{C}^+$:
\begin{align}
    \mathcal{C^+}:\quad\cos{\tilde{\Psi}^+}=\frac{\sqrt{\pa{3+\cos^2{\tilde{\Upsilon}^+}}\cos^2{\tilde{\Upsilon}^+}}-2}{4+\cos^2{\tilde{\Upsilon}^+}}.
\end{align}
Thus, the critical curve for a prograde NHEK orbiter is
\begin{align}
    \label{eq:CriticalCurveNHEK}
	\cos{\tilde{\Psi}}=
	\begin{cases}
        \frac{\sqrt{\pa{3+\cos^2{\tilde{\Upsilon}}}\cos^2{\tilde{\Upsilon}}}-2}{4+\cos^2{\tilde{\Upsilon}}}
        &0\le\tilde{\Upsilon}\le\pi,\\
	    0&-\pi\le\tilde{\Upsilon}\le0.
	\end{cases}
\end{align}
We can also calculate the redshift $\tilde{g}$ of critical photons.  For a NHEK orbiter \eqref{eq:DoubleLimit}, the redshift \eqref{eq:RedshiftPsi} reduces to
\begin{align}
    \label{eq:CriticalRedshiftNHEK}
    g=\frac{1}{\sqrt{3}}+\frac{2}{\sqrt{3}}\cos{\tilde{\Psi}}
\end{align}
in the $\kappa\to0$ limit, with $\cos{\tilde{\Psi}}$ given by Eq.~\eqref{eq:CriticalCurveNHEK}, while
\begin{align}
    \label{eq:CriticalXi}
    \frac{\xi_s}{\sqrt{M\Delta(r_s)}}\to\frac{\sqrt{3}}{2}.
\end{align}

In the next sections, we will use the shape of the critical curve in the sky of an isotropically emitting orbiter to compute properties of its emission, such as the photon escape probability and total flux radiated to infinity.

\section{Escape probability}
\label{sec:EscapeProbability}

\begin{figure*}
    \centering
	\includegraphics[width=\columnwidth]{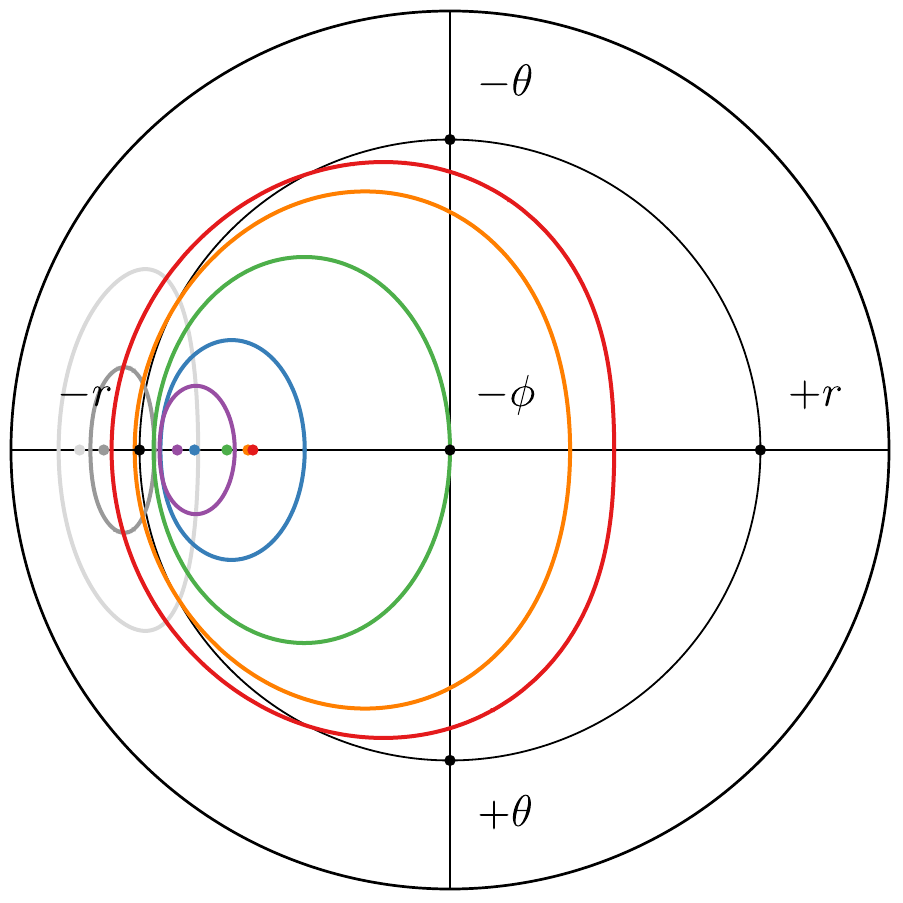}
	\includegraphics[width=\columnwidth]{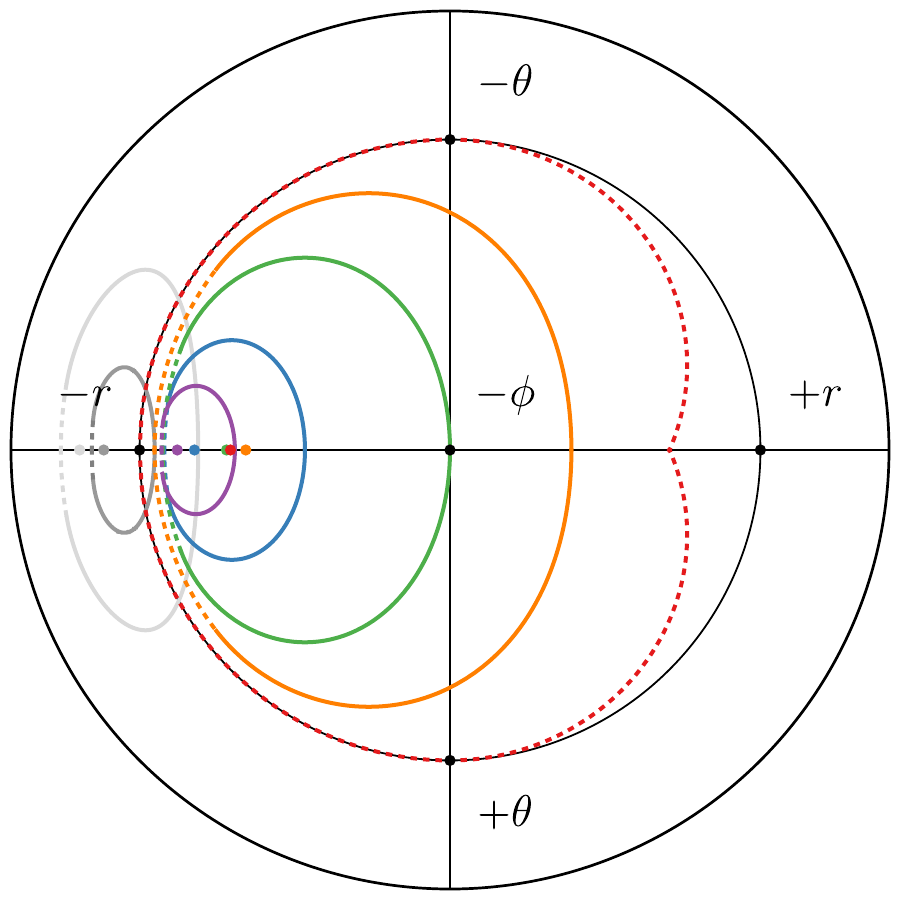}
	\caption{Critical curve in the backside projection \eqref{eq:BacksideProjection} of the circular equatorial orbiter sky.  The projection is area-preserving, so each of the eight octants has the same area (and solid angle) $\pi/2$.  The outer circle is a point (the $+\phi$ direction).  We show near-extremal BH spin $a/M=99\%$ (left) and extremal spin $a=M$ (right).  The orbital radii $r_s$ are $20M$ (purple), $10M$ (blue), $\rt_+\approx4M$ (outermost photon shell, green), $2M$ (ergosphere, orange), and $r_\mathrm{ms}^+$ (red) for prograde orbiters, and $20M$ (gray) and $r_\mathrm{ms}^-$ (light gray) for retrograde orbiters.  The colored dots indicate the corresponding ``direction to the BH center'' [Eq.~\eqref{eq:BlackHoleCenter}].  As $r_s\to\infty$, the critical curve shrinks to that point (the $-r$ direction).  It touches the vertical axis at $r_s=\rt_+$, when the orbiter enters the photon shell.  For nonextremal BH spins, the critical curve is parametrized by photon shell radius $\rt$ [Eq.~\eqref{eq:BacksideCriticalCurve}].  At extremality, this parametrization only covers the part of the photon shell that remains outside the throat (solid curves), but in order to close, the full critical curve also requires contributions from photon shell radii $\Rt$ inside of it (dotted curves) [Eqs.~\eqref{eq:BacksideCriticalCurveExtremeKerr}].  If the orbiter itself lies in the NHEK geometry, then the entire curve corresponds to photon shell radii in the throat [Eq.~\eqref{eq:BacksideCriticalCurveNHEK}].}
	\label{fig:BacksideProjection}
\end{figure*}

\begin{figure}
    \centering
	\includegraphics[width=\columnwidth]{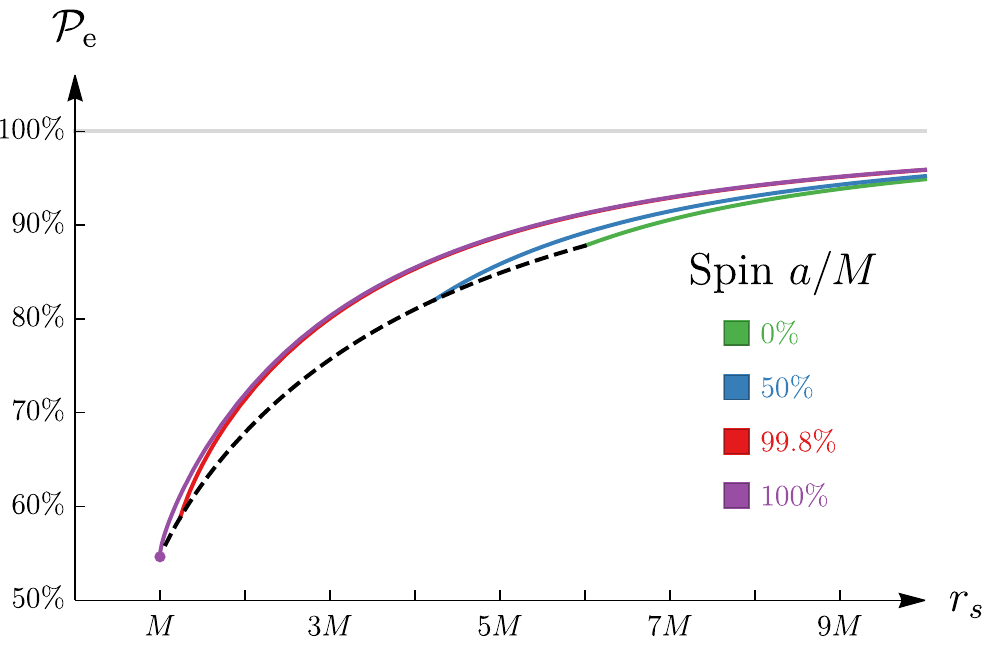}
	\caption{Escape probability from a prograde emitter as a function of orbital radius $r_\mathrm{ms}^+\le r_s\le10M$ for fixed values of BH spin.  The black dashed curve is the escape probability from the ISCO as a function of BH spin and defines the envelope of the other curves.  The plots for retrograde emitters look similar and are not shown.  When $a=M$, the general formula \eqref{eq:CaptureArea} breaks down and gives a manifestly wrong answer, which misses a substantial contribution from the NHEK geometry [Eq.~\eqref{eq:CaptureAreaExtremeKerr}].  The purple dot corresponds to the critical fixed point \eqref{eq:EscapeProbabilityNHEK} of the photon escape probability from the throat.}
	\label{fig:EscapeProbability}
\end{figure}

From now on, we consider circular equatorial orbiters that emit photons isotropically in their rest frame.  In this section, we use the shape of the critical curve in the orbiter sky that we derived in Sec.~\ref{sec:SkyCriticalCurve} to compute the photon escape probability as a function of BH spin and orbital radius.  The probability that a photon escapes to infinity (rather than being captured by the BH) is given by the fraction of solid angle in the emitter sky corresponding to directions of escape,
\begin{align}
    \label{eq:EscapeProbability}
    \mathcal{P}_\mathrm{e}=1-\frac{\mathcal{A}}{4\pi},
\end{align}
where $\mathcal{A}$ is the area of the BH capture region in the sky.

To compute this escape probability, we first define an area-preserving projection of the emitter sky to the plane.  Such a projection requires the area element on the plane $\ed\mathcal{A}$ and the area element on the sphere $\ed\Omega$ to be equal:
\begin{align}
    \ed\mathcal{A}=\rho\ed\rho\wedge\ed\varphi
    =\sin{\Psi}\ed\Psi\wedge\ed\Upsilon
	=\ed\Omega.
\end{align}
We choose to employ a backside projection
\begin{align}
    \rho\ed\rho=-\sin{\Psi}\ed\Psi,\quad
    \ed\varphi=-\ed\Upsilon,
\end{align}
with $\left.\rho\right|_{\Psi=\pi}=0$ and $\left.\varphi\right|_{\Upsilon=0}=\pi/2$, so that
\begin{align}
    \label{eq:BacksideProjection}
    \rho=2\cos\pa{\frac{\Psi}{2}},\quad
    \varphi=\frac{\pi}{2}-\Upsilon.
\end{align}
This amounts to looking at the sphere of the emitter sky in Fig.~\ref{fig:Angles} from the outside, projecting it from the left ($-\phi$ direction) onto the plane separating its forward and backward hemispheres.  This area-preserving backside projection is illustrated in Fig.~\ref{fig:BacksideProjection}.  Its center $\rho=0$ is the $-\phi$ direction on the sphere and its vertical axis is aligned with the local zenith ($-\theta$ direction).  The sphere is unfolded onto the plane such that the backward hemisphere is mapped to the disk $0\le\rho\le\sqrt{2}$ and the forward hemisphere to the annulus $\sqrt{2}\le\rho\le2$.  The angle $\varphi$ increases from $\varphi=0$ in the $+r$ direction ($\Upsilon=\pi/2$) to $\varphi=\pi/2$ in the $-\theta$ direction ($\Upsilon=0$), growing in the opposite sense to the azimuthal angle $\Upsilon$, which is defined in the same plane.  This projection is inspired by, but different from, the one defined by Cunningham and Bardeen \cite{Cunningham1973} [Eq.~(37)] and illustrated in Fig.~3 therein (they chose to project the northern hemisphere down from the local zenith $\Theta=0$ into the equatorial plane).

Under the backside projection \eqref{eq:BacksideProjection}, the orbiter sky critical curve \eqref{eq:SkyCriticalCurve} is mapped to the closed plane curve
\begin{align}
    \label{eq:BacksideCriticalCurve}
    \mathcal{C}=\cu{\pa{\tilde{\rho}(\rt),\tilde{\varphi}(\rt)}\Big|\rt_-\le\rt\le\rt_+}.
\end{align}
In Fig.~\ref{fig:BacksideProjection}, we plot a number of these projected critical curves for several orbital radii and two BH spins: the near-extremal spin $a/M=99\%$ and the extremal spin $a=M$.  As usual, the latter case requires special care.

The plots in Fig.~\ref{fig:BacksideProjection} render manifest a useful feature of the backside projection \eqref{eq:BacksideProjection} that guided us to it: it always maps the region of photon capture in the emitter sky to the interior $\mathcal{C}^\circ$ of the projected critical curve \eqref{eq:BacksideCriticalCurve}, which always contains the direction to the black hole \eqref{eq:BlackHoleCenter}.\footnote{For prograde orbiters, $\mathcal{C}^\circ$ may include the $\Psi=\pi$ backward direction (the center of the projection), but never the $\Psi=0$  forward direction (its outer circle).  For retrograde orbiters, it never includes either of these directions.  Hence, photons emitted in the direction of orbital motion always escape, as claimed in Sec.~\ref{sec:SkyCriticalCurve} above.}  As such, the area $\mathcal{A}$ of the escape region in the emitter sky is that of the critical curve's interior $\mathcal{C}^\circ$,
\begin{align}
    \label{eq:CaptureArea}
    \mathcal{A}&=\int_{\mathcal{C}^\circ}\ed\mathcal{A}
    =\int_{\mathcal{C}^\circ}\rho\ed\rho\ed\varphi
    =\int_{\rt_-}^{\rt_+}\tilde{\rho}^2(\rt)\frac{d\tilde{\varphi}(\rt)}{d\rt}\ed\rt,
\end{align}
where the $1/2$ from the $\ed\pa{\rho^2/2}$ integral canceled against a factor of $2$ accounting for the fact that, for each choice of sign $\pm_\theta$, the $\rt$ parametrization only covers half of the (reflection-symmetric) critical curve.  In order to obtain the correct (positive) area, we must ensure that
\begin{align}
    \label{eq:IntegrationSign}
    \frac{d\tilde{\varphi}(\rt_-)}{d\rt}>0,
\end{align}
which requires choosing the sign $\pm_\theta=+$ in $\tilde{\varphi}=\pi/2-\tilde{\Upsilon}$.  This amounts to integrating only over the lower half of the projection plane $-\pi<\varphi\le0$ in Fig.~\ref{fig:BacksideProjection}.

For generic values of the BH spin $0\le a<M$, the projected critical curve \eqref{eq:BacksideCriticalCurve} has a complicated parametrization and the integral \eqref{eq:CaptureArea} is analytically intractable.  On the other hand, we will now show that simplifications occur in the (near-)extremal regime $a\to M$.  In particular, for emitters in the throat, the escape probability tends to a critical fixed point with finite escape probability.  We plot the photon escape probability in Fig.~\ref{fig:EscapeProbability}.

\subsection{Extreme Kerr emitter}

The critical curve of an extreme Kerr emitter \eqref{eq:CriticalCurveExtremeKerr} is mapped under the backside projection \eqref{eq:BacksideProjection} to
\begin{subequations}
\label{eq:BacksideCriticalCurveExtremeKerr}
\begin{align}
    \mathcal{C}&=\mathcal{C}_f\cup\mathcal{C}_n,\\
    \mathcal{C}_f&=\cu{\pa{\tilde{\rho}(\rt),\tilde{\varphi}(\rt)}\Big|M\le\rt\le 4M},\\
    \mathcal{C}_n&=\cu{\pa{\tilde{\rho}_n,\tilde{\varphi}}\Big|\pi-\tilde{\varphi}_n\leq\tilde{\varphi}\leq\pi+\tilde{\varphi}_n},
\end{align}
\end{subequations}
where
\begin{align}
    \tilde{\rho}_n^2&=\frac{2(r_s+2M)}{r_s\pm\sqrt{Mr_s}+M},\\
    \label{eq:ExtremeCurveAngle}
    \tilde{\varphi}_n&=\arcsin{\sqrt{\frac{3M^2}{r_s(r_s+2M)}}}.
\end{align}
The curve $\mathcal{C}_n$ is the generalization to finite orbital radius of the NHEKline \cite{Gralla2017}.

Using the formula \eqref{eq:BacksideCriticalCurve} (with $\pm_\theta=+$ to ensure the condition \eqref{eq:IntegrationSign} is satisfied), the capture area is given by
\begin{align}
    \label{eq:CaptureAreaExtremeKerr}
    \mathcal{A}&=\int_M^{4M}\tilde{\rho}^2(\rt)\frac{d\tilde{\varphi}(\rt)}{d\rt}\ed\rt+\tilde{\rho}_n^2\tilde{\varphi}_n.
\end{align}
Plugging this into Eq.~\eqref{eq:EscapeProbability} yields the photon escape probability for an extreme Kerr emitter, which we plot as a function of orbital radius $r_s$ in Fig.~\ref{fig:EscapeProbability} (purple curve).

\subsection{NHEK emitter}

By Eqs.~\eqref{eq:CriticalCurveNHEK} and \eqref{eq:BacksideProjection}, the critical curve in the backside projection of a NHEK emitter becomes
\begin{align}
    \label{eq:BacksideCriticalCurveNHEK}
    \tilde{\rho}&=
    \begin{cases}
        \sqrt{2}&\frac{\pi}{2}\le\tilde{\varphi}\le\frac{3 \pi}{2},\\
        \sqrt{2+\frac{2\sqrt{\sin^2{\tilde{\varphi}}\pa{3+\sin^2{\tilde{\varphi}}}}-4}{4+\sin^2{\tilde{\varphi}}}}&-\frac{\pi}{2}\le\tilde{\varphi}\le\frac{\pi}{2}.
    \end{cases}
\end{align}
Using this simple formula, the capture area reduces to
\begin{align}
     \mathcal{A}&=\pi+\int_{\frac{\pi}{2}}^\pi\tilde{\rho}^2(\tilde{\varphi})\ed\tilde{\varphi}
     =\frac{7\pi}{3}-\frac{4\arctan\sqrt{5/3}}{\sqrt{5}}.
\end{align}
Therefore, the escape probability \eqref{eq:EscapeProbability} for a NHEK emitter in the throat tends to the (nonzero) critical fixed point 
\begin{align}
    \label{eq:EscapeProbabilityNHEK}
    \mathcal{P}_\mathrm{e}=\frac{5}{12}+\frac{\arctan\sqrt{5/3}}{\sqrt{5}\pi}
    \approx54.6455\%,
\end{align}
illustrated with a dot in Fig.~\ref{fig:EscapeProbability}.  This exact formula agrees with the numerical value recently computed in Ref.~\cite{Igata2020}, and Fig.~5 therein is consistent with our 
Fig.~\ref{fig:EscapeProbability}.

\section{Flux to celestial sphere}
\label{sec:FluxCurves}

\begin{figure*}
    \centering
	\includegraphics[width=\textwidth]{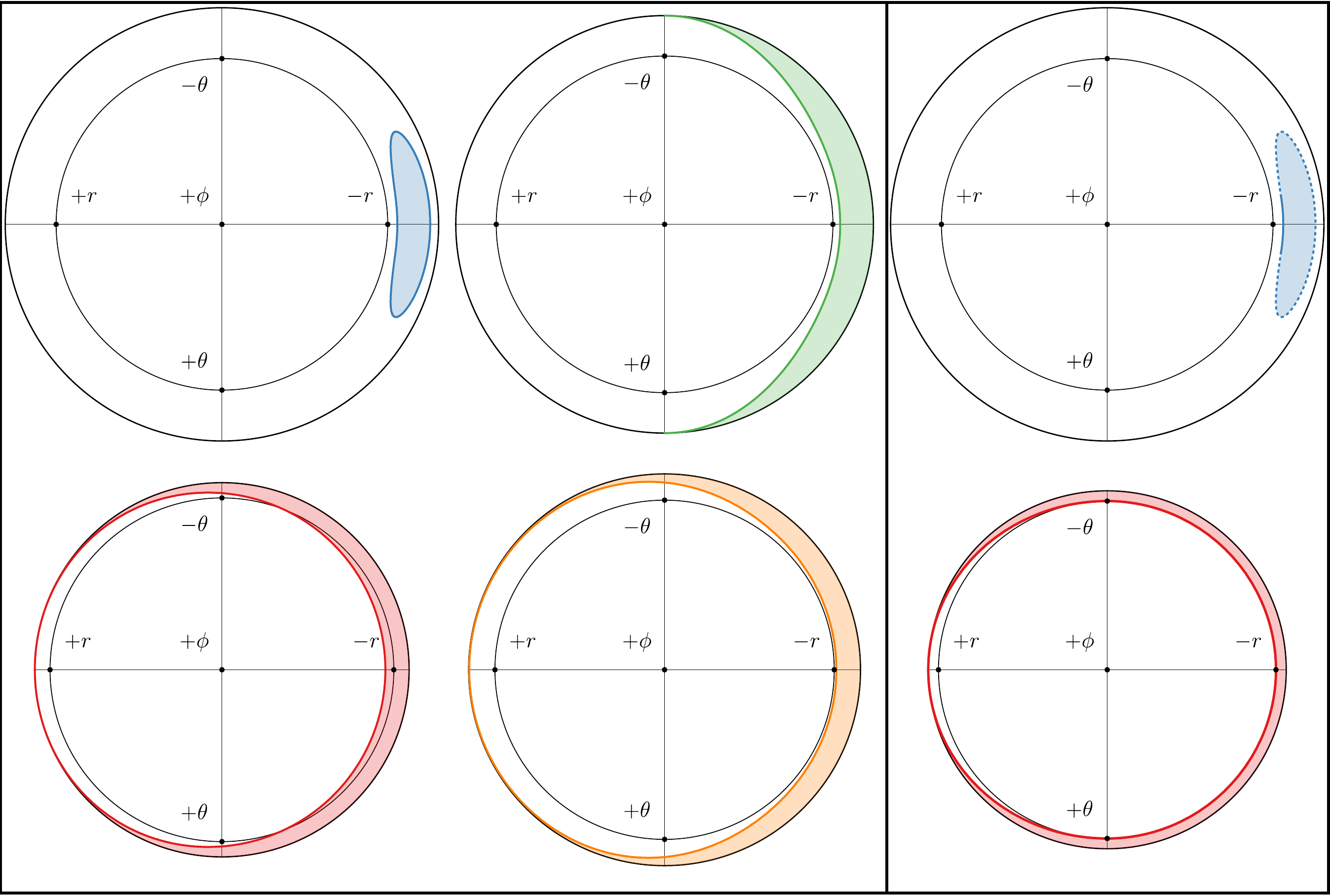}
	\caption{Critical curve in the redshift-weighted projection \eqref{eq:FrontsideProjection}.  Area on the projection corresponds to time-averaged flux received at infinity, not area in the emitter sky (Fig.~\ref{fig:BacksideProjection}).  Directions corresponding to light rays with formally negative redshift, which cannot reach infinity, are excluded, so the outer circle is either a point (the $-\phi$ direction) or a circle of fixed $\Psi=\Psi_\mathrm{max}$ [Eq.~\eqref{eq:PsiMax}].  We show near-extremal BH spin $a/M=99.8\%$ (left and middle columns) and extremal spin $a=M$ (right column).  The orbital radii $r_s$ are $10M$ (blue), $\rt_+\approx4M$ (outermost photon shell, green), $2M$ (ergosphere, orange), and $r_\mathrm{ms}^+$ (red) for prograde orbiters only, with coloring to indicate the region of BH capture.  As $r_s\to\infty$, the critical curve shrinks to a point near the $-r$ direction.  It touches the vertical axis at $r_s=\rt_+$, when the orbiter enters the photon shell.  For nonextremal BH spins, the critical curve is parametrized by photon shell radius $\rt$ [Eq.~\eqref{eq:FrontsideCriticalCurve}].  At extremality (upper right panel), this parametrization only covers the part of the photon shell that remains outside the throat (dotted curve), but in order to close, the full critical curve also requires contributions from photon shell radii $\Rt$ inside of it (solid curve) [Eqs.~\eqref{eq:FrontsideCriticalCurveExtremeKerr}].  If the orbiter itself lies in the NHEK geometry (lower right panel), then the entire curve is obtained by photon shell radii in the throat [Eq.~\eqref{eq:FrontsideCriticalCurveNHEK}].}
	\label{fig:RedshiftProjection}
\end{figure*}

\begin{figure}
    \centering
	\includegraphics[width=\columnwidth]{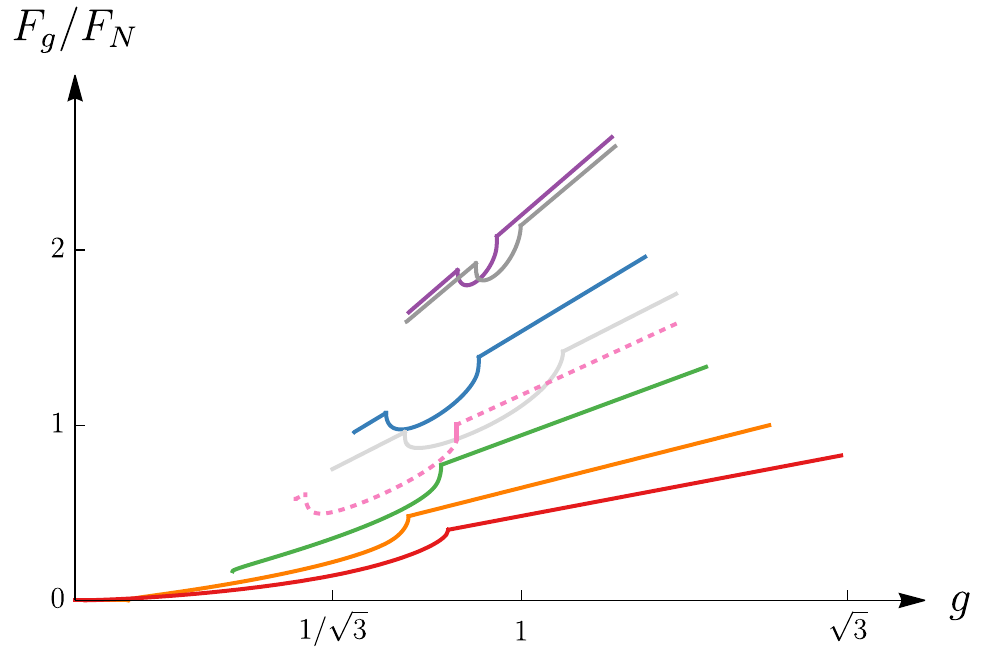}
	\caption{Total flux $F_g$ collected on the celestial sphere as a function of observed redshift, normalized by Newtonian flux $F_N$ [Eq.~\eqref{eq:FluxCurve}].  The spin is $a/M=99.8\%$ and the orbital radii $r_s$ are $20M$ (purple), $10M$ (blue), $\rt_+\approx4M$ (outermost photon shell, green), $2M$ (ergosphere, orange), and $r_\mathrm{ms}^+$ (red) for prograde orbiters, and $20M$ (gray) and $r_\mathrm{ms}^-$ (light gray) for retrograde orbiters.  We also display a prograde emitter at radius $r_s=6M$ orbiting an extremal BH with $a=M$ (pink curve).  All orbiters emit the same monochromatic and isotrophic flux $F_N$ in their rest frame, so the relative magnitude of these curves is physical.  As $r_s\to\infty$, the line is unbroadened and $F_g$ tends to a delta-function spike $F_N\delta(g-1)$.  The curves are composed of either two or three segments [Eq.~\eqref{eq:EscapeAngle}], and are mostly linear except for a ``bite'' corresponding to photon capture by the BH and whose shape is given by the critical curve [Eq.~\eqref{eq:Bite}].  For extreme Kerr orbiters, $F_g$ develops a discontinuity (solid pink line) at $g=\tilde{g}_-$ [Eq.~\eqref{eq:ExtremeFluxNHEKline}].  This vertical line is the spectral manifestation of the NHEKline.}
	\label{fig:FluxCurves}
\end{figure}

Due to gravitational redshift and Doppler shift, the radiation emitted by a monochromatic orbiter is spectrally broadened at infinity in a characteristic spin-dependent way (see, e.g., Ref.~\cite{Gates2020} for a recent discussion).  In this section, we ask: how does the total spectral flux received at infinity (the flux per unit frequency, integrated over the celestial sphere) vary with observed redshift?  We obtain a remarkably simple answer by introducing a redshift-weighted projection of the emitter sky, such that equal areas on the projection plane correspond to equal amounts of energy flux collected on the celestial sphere.  Such a projection requires the area element on the plane $\ed\mathcal{A}_g$ to equal the area element on the sphere $\ed\Omega$ times the observed redshift of the corresponding light rays:
\begin{align}
    \ed\mathcal{A}_g=\rho_g\ed\rho_g\wedge\ed\varphi_g
    =g(\Psi)\sin{\Psi}\ed\Psi\wedge\ed\Upsilon
	=g\ed\Omega.
\end{align}
We choose to employ a frontside projection
\begin{align}
    \rho_g\ed\rho_g=g(\Psi)\sin{\Psi}\ed\Psi,\quad
    \ed\varphi_g=\ed\Upsilon,
\end{align}
with $\left.\rho_g\right|_{\Psi=0}=0$ and $\left.\varphi_g\right|_{\Upsilon=0}=\pi/2$.  After integrating $-g(\cos{\Psi})\ed\cos{\Psi}$ using Eq.~\eqref{eq:RedshiftPsi}, this results in
\begin{subequations}
\label{eq:FrontsideProjection}
\begin{align}
    \label{eq:RedshiftRadius}
    \rho_g&=2\sin\pa{\frac{\Psi}{2}}\sqrt{g(\Psi)\pm\frac{\sqrt{M\Delta(r_s)}}{\xi_s}\sin^2\pa{\frac{\Psi}{2}}},\\
    \label{eq:RedshiftAngle}
    \varphi_g&=\frac{\pi}{2}+\Upsilon.
\end{align}
\end{subequations}
This amounts to looking at the sphere of the emitter sky in Fig.~\ref{fig:Angles} from the outside, projecting it from the right ($+\phi$ direction) onto the plane separating its forward and backward hemispheres.  This redshift-weighted projection is illustrated in Fig.~\ref{fig:RedshiftProjection}.  Its center $\rho=0$ is the $+\phi$ direction on the sphere and its vertical axis is aligned with the local zenith ($-\theta$ direction).  The sphere is unfolded onto the plane such that the forward hemisphere is mapped to the disk $0\le\rho_g\le\rho_g(\pi/2)$ and the backward hemisphere to the annulus $\rho_g(\pi/2)\le\rho_g\le\rho_g(\Psi_\mathrm{max})$, with $\Psi_\mathrm{max}$ given in Eq.~\eqref{eq:PsiMax}.  The angle $\varphi_g$ increases from $\varphi_g=0$ in the $-r$ direction ($\Upsilon=-\pi/2$) to $\varphi_g=\pi/2$ in the $-\theta$ direction ($\Upsilon=0$), growing in the same sense as the azimuthal angle $\Upsilon$, which is defined in the same plane.  Directions in the emitter sky with negative redshift---for which photons are necessarily captured by the BH---are not included in this projection.  This projection is almost identical to that defined by Cunningham and Bardeen \cite{Cunningham1973} [Eqs.~(38)--(39)] and illustrated in Fig.~4 therein, except that we chose to glue the projections of the two hemispheres by extending the forward disk with the backward annulus, rather than have a separate backward disk.

Under the redshift-weighted projection \eqref{eq:FrontsideProjection}, the emitter sky critical curve is mapped to the closed plane curve
\begin{align}
    \label{eq:FrontsideCriticalCurve}
    \mathcal{C}=\cu{\pa{\tilde{\rho}_g(\rt),\tilde{\varphi}_g(\rt)}\Big|\rt_-\le\rt\le\rt_+}.
\end{align}
In Fig.~\ref{fig:RedshiftProjection}, we plot the critical curve in this redshift-weighted projection for BH spin $a/M=99.8\%$ and several orbiter radii.  A special feature of this curve is that the total flux received on the celestial sphere is proportional to the area integral over its interior $\mathcal{C}^\circ$, which now corresponds to the escape region $\mathcal{E}$:
\begin{align}
    F&=\frac{F_N}{4\pi}\int_\mathcal{E}\ed\mathcal{A}_g
    =\frac{F_N}{4\pi}\int_\mathcal{E}\rho_g\ed\rho_g\ed\varphi_g\\
    &=\frac{F_N}{4\pi}\int_\mathcal{E}\rho_g(g)\ab{\frac{d\rho_g(g)}{dg}}\ed g\ed\varphi_g.
\end{align}
The normalization factor $F_N$ is the Newtonian flux that would be collected on the celestial sphere if the emitter were in flat spacetime.  Inverting Eq.~\eqref{eq:RedshiftPsi} for $\Psi(g)$ yields
\begin{align}
    \cos{\Psi}=\pm\frac{g\xi_s-\pa{r_s^{3/2}-2M\sqrt{r_s}\pm a\sqrt{M}}}{\sqrt{M\Delta(r_s)}}.
\end{align}
Plugging this into Eq.~\eqref{eq:RedshiftRadius} defines $\rho_g(g)$, which obeys the remarkably simple formula
\begin{align}
    \rho_g(g)\frac{d\rho_g(g)}{dg}=\mp\frac{g\xi_s}{\sqrt{M\Delta(r_s)}}.
\end{align}
Hence,
\begin{align}
    F=\frac{F_N}{4\pi}\frac{\xi_s}{\sqrt{M\Delta(r_s)}}\int_\mathcal{E}g\ed g\ed\varphi
    =\ab{\int_{\hat{g}}^{g_0}F_g\ed g},
\end{align}
where in the last step, we introduced the total flux $F_g$ received on the celestial sphere with redshift $g$,
\begin{align}
    \label{eq:FluxCurve}
    \frac{F_g}{F_N}=\frac{1}{4\pi}\frac{g\xi_s}{\sqrt{M\Delta(r_s)}}\int_{-\varphi_\mathcal{E}(g)}^{\varphi_\mathcal{E}(g)}\ed\varphi
    =\frac{1}{2\pi}\frac{\xi_sg\varphi_\mathcal{E}(g)}{\sqrt{M\Delta(r_s)}},
\end{align}
with the ``escape angle'' $\varphi_\mathcal{E}(g)$ defined as half the opening angle of the arc of the circle $\rho(g)$ that lies inside $\mathcal{E}$ (Fig.~\ref{fig:RedshiftProjection}), which corresponds to emission received at infinity with redshift $\hat{g}\lessgtr g\lessgtr g_0$, with the upper/lower inequalities according to whether the orbit is prograde/retrograde, respectively.  This escape angle is given by
\begin{align}
    \label{eq:EscapeAngle}
    \varphi_\mathcal{E}(g)=
    \begin{cases}
        \pi-\tilde{\varphi}_g(\rt_+)&\hat{g}\lessgtr g\lessgtr\tilde{g}_+,\\
        \pi-\tilde{\varphi}_g(\rt(g))&\tilde{g}_+\lessgtr g\lessgtr\tilde{g}_-,\\
        \pi&\tilde{g}_-\lessgtr g\le g_0,
    \end{cases}
\end{align}
where $\rt(g)$ is the inverse of $\tilde{g}(\rt)$ given by Eq.~\eqref{eq:CriticalRedshift}, and we must choose $\pm_\theta=-$ to get the correct contribution.  Recall from Eq.~\eqref{eq:RedshiftBounds} that we always have the ordering $\tilde{g}_+\lessgtr\tilde{g}_-\lessgtr g_0$, but prograde orbiters may have $\hat{g}=\tilde{g}_+$, in which case the first interval does not exist.

In Fig.~\ref{fig:FluxCurves}, we plot $F_g/F_N$ for emitters orbiting a BH of near-extremal spin $a/M=99.8\%$ at several orbital radii.  This function is mostly linear, $F_g\propto g$, except for the interval $\tilde{g}_+\lessgtr g\lessgtr\tilde{g}_-$, in which a portion of the flux is ``bitten'' out of the curve because of photon capture by the BH.  This BH ``bite'' is the spectral signature of the critical curve $\mathcal{C}$, and indeed we use $\tilde{g}(\rt)$ to plot it using its parametrization by photon shell radius,
\begin{align}
    \label{eq:Bite}
    \mathcal{C}_\mathrm{bite}=\cu{\left.\pa{\tilde{g},\frac{1}{2\pi}\frac{\xi_s\tilde{g}\varphi_\mathcal{E}\pa{\tilde{g}}}{\sqrt{M\Delta(r_s)}}}\right|\rt_-\le\rt\le\rt_+}.
\end{align}

\subsection{Extreme Kerr emitter}

The critical curve of an extreme Kerr emitter \eqref{eq:CriticalCurveExtremeKerr} is mapped under the redshift-weighted projection \eqref{eq:FrontsideProjection} to
\begin{subequations}
\label{eq:FrontsideCriticalCurveExtremeKerr}
\begin{align}
    \mathcal{C}&=\mathcal{C}_f\cup\mathcal{C}_n,\\
    \mathcal{C}_f&=\cu{\pa{\tilde{\rho}_g(\rt),\tilde{\varphi}_g(\rt)}\Big|M\le\rt\le4M},\\
    \mathcal{C}_n&=\cu{\pa{\tilde{\rho}_{g,n},\tilde{\varphi}_g}\Big|-\tilde{\varphi}_n\leq\tilde{\varphi}_g\leq\tilde{\varphi}_n},
\end{align}
\end{subequations}
where $\tilde{\varphi}_n$ is the same as in Eq.~\eqref{eq:ExtremeCurveAngle}, while
\begin{align}
    \tilde{\rho}_{g,n}^2=\frac{\pa{r_s\pm2\sqrt{Mr_s}}^{3/2}\br{2r_s+M\pm\sqrt{Mr_s}}}{\sqrt{r_s}\br{r_s\pm\sqrt{Mr_s}+M}^2}.
\end{align}
The curve $\mathcal{C}_n$ is the generalization to finite orbital radius of the NHEKline \cite{Gralla2017}, which in this projection is not a straight line but rather a circular arc corresponding to
\begin{align}
    \label{eq:ExtremeFluxNHEKline}
    g=\tilde{g}_-
    =\frac{\sqrt{r_s\pa{r_s^2-3Mr_s\pm2M\sqrt{Mr_s}}}}{r_s^{3/2}\mp M^{3/2}}.
\end{align}
This function monotonically decreases from $g=1$ when $r_s\to\infty$ to $g=1/\sqrt{3}$ for a prograde orbit with $r_\mathrm{ms}^+\to M$, or $g=3\sqrt{3}/7$ for a retrograde orbit at $r_\mathrm{ms}^-=9M$.

The flux curve $F_g$ can be computed for extreme Kerr orbiters using Eq.~\eqref{eq:FluxCurve}, with the escape angle still given by Eq.~\eqref{eq:EscapeAngle}.  There is, however, a new feature: the NHEKline $\mathcal{C}_n$ manifests itself as a vertical line in the flux curve, creating a discontinuity at $g=\tilde{g}_-$ [Eq.~\eqref{eq:ExtremeFluxNHEKline}].  We display such a curve in Fig.~\ref{fig:FluxCurves} (pink curve).

\subsection{NHEK emitter}

\begin{figure}
    \centering
	\includegraphics[width=\columnwidth]{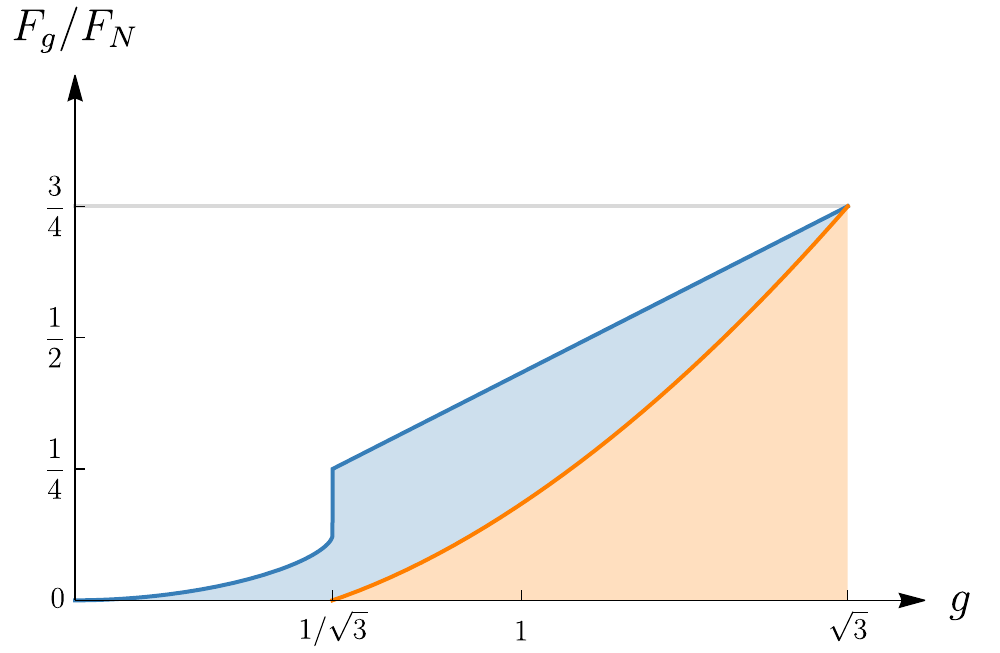}
	\caption{Blue: total flux $F_g$ received on the celestial sphere from a NHEK orbiter, as a function of observed redshift and normalized by Newtonian flux $F_N$ [Eq.~\eqref{eq:CriticalFlux}].  Orange: same quantity obtained by integrating an approximate formula for the angle-dependent flux $F_g(\theta_o)$ over the celestial sphere [Eq.~\eqref{eq:ApproximateCriticalFlux}].  The approximation can account for $60\%+$ of the exact flux and reproduces the correct (linear) behavior at high energy (blueshifts $g\to\sqrt{3}$), but misses significant contributions to the low-energy spectrum (redshifts $g\lesssim1/\sqrt{3}$).}
	\label{fig:CriticalFluxCurve}
\end{figure}

Using Eqs.~\eqref{eq:CriticalCurveNHEK}--\eqref{eq:CriticalXi} and \eqref{eq:FrontsideProjection}, the critical curve in the redshift-weighted projection of a NHEK emitter becomes
\begin{align}
    \label{eq:FrontsideCriticalCurveNHEK}
    \tilde{\rho}_g(\tilde{\varphi})=
    \begin{cases}
        X\pa{\sin^2{\tilde{\varphi}}}
        &\frac{\pi}{2}\le\tilde{\varphi}\le\frac{3\pi}{2},\\
    	\frac{2}{3^{1/4}}
    	&-\frac{\pi}{2}\le\tilde{\varphi}\le\frac{\pi}{2},
	\end{cases}
\end{align}
where we introduced a function $X(s)$ whose square is
\begin{align}
    \br{X(s)}^2=\frac{2}{\sqrt{3}}\frac{s^2+15s+36-s\sqrt{s(3+s)}}{\pa{4+s}^2}.
\end{align}
Note that $\tilde{\rho}_g(\tilde{\varphi})$ lies in the interval
\begin{align}
    \frac{3^{3/4}}{\sqrt{2}}\le\tilde{\rho}_g(\tilde{\varphi})\le\frac{2}{3^{1/4}},
\end{align}
whose lower radius corresponds to the projection of the equator dividing the forward and backward hemispheres, where the emission has $g=1/\sqrt{3}$.

For NHEK emitters, we can invert Eqs.~\eqref{eq:CriticalCurveNHEK}--\eqref{eq:CriticalRedshiftNHEK} as
\begin{align}
    \cos^2{\tilde{\Upsilon}}=\frac{\pa{1+2\cos{\tilde{\Psi}}}^2}{1-\cos^2{\tilde{\Psi}}},\quad
    \cos{\Psi}=\frac{\sqrt{3}g-1}{2}.
\end{align}
Combining these equations yields
\begin{align}
    \tilde{\Upsilon}_g(g)=\frac{12g^2}{3+2\sqrt{3}g-3g^2}.
\end{align}
By Eq.~\eqref{eq:RedshiftAngle}, the escape angle \eqref{eq:EscapeAngle} simplifies to
\begin{align}
    \varphi_\mathcal{E}(g)=
    \begin{cases}
        \frac{\pi}{2}-\tilde{\Upsilon}_g(g)
        &0\le g<\frac{1}{\sqrt{3}},\\
        \pi
        &\frac{1}{\sqrt{3}}<g\leq\sqrt{3}.
    \end{cases}
\end{align}
Since $F_g/F_N=\sqrt{3}g\varphi_\mathcal{E}(g)/(4\pi)$, as can be seen by plugging Eq.~\eqref{eq:CriticalXi} into Eq.~\eqref{eq:FluxCurve}, it follows that
\begin{align}
    \label{eq:CriticalFlux}
    \frac{F_g}{F_N}=
    \begin{cases}
        \frac{\sqrt{3}}{4\pi}g\arcsin\pa{\frac{2g}{\sqrt{1+2g/\sqrt{3}-g^2}}}
        &0\le g<\frac{1}{\sqrt{3}},\\
        \frac{\sqrt{3}}{4}g
        &\frac{1}{\sqrt{3}}<g\leq\sqrt{3}.
    \end{cases}
\end{align}
This flux curve is the same for all NHEK emitters, regardless of their depth in the throat, and can therefore be regarded as another example of universal critical behavior of a high-spin Kerr BH.  This formula is one of our main new results, and we plot it in Fig.~\ref{fig:CriticalFluxCurve} (blue curve).  It implies that orbiters continue to emit finite flux to infinity even as $r_s\to M$, in apparent contradiction with Fig.~6 of Ref.~\cite{Bardeen1973}, which preceded the discovery of the NHEK region.

As in the case of the extreme Kerr emitter, the escape angle $\varphi_\mathcal{E}(g)$ and the flux $F_g$ both present a discontinuity at $g=\tilde{g}_-=1/\sqrt{3}$, which is the manifestation of the NHEKline portion of the critical curve in the BH bite.  Note that for the NHEK emitter,
\begin{align}
    \label{eq:Discontinuity}
    \lim_{g\to\pa{1/\sqrt{3}}^+}F_g=2\lim_{g\to\pa{1/\sqrt{3}}^-}F_g.
\end{align}
The factor of 2 has a simple interpretation discussed in Sec.~\ref{sec:Comparison} below, which can be visually seen in Fig.~\ref{fig:DiskEnvelope}. 

\subsection{Thin equatorial disk}
\label{sec:NHEKDiskTotalFlux}

In general, the total flux received at infinity with observed redshift $g$ from an entire thin disk of circular equatorial orbiters emitting monochromatically is
\begin{align}
    F_\mathrm{disk}(g)&=\int\frac{F_g}{F_N}(r_s)J(r_s)\ed r_s\ed\phi_s\\
    \label{eq:FluxDisk}
    &=\int\frac{\xi_sg\varphi_\mathcal{E}(g)}{\sqrt{M\Delta(r_s)}}J(r_s)\ed r_s,
\end{align}
where the radial profile $J(r_s)$ is a local surface emissivity in the orbiters' frame, and we are effectively assuming that the disk is stationary and axisymmetric, since $F_g(r_s)$ is the collected flux time-averaged over an orbit.

Our result \eqref{eq:CriticalFlux} shows that an observer receives finite flux from every emission radius in NHEK.  Hence, the portion of the disk that lies in the throat contributes
\begin{align}
    \label{eq:FluxDiskNHEK}
    F_\mathrm{disk}^\mathrm{NHEK}(g)=\frac{F_g}{F_N}\int_\mathrm{NHEK}J(r_s)\ed r_s
\end{align}
to the total flux \eqref{eq:FluxDisk}.  Note that the dependence on the observed redshift $g$ separates from the radial integral over the emissivity, which only provides an overall constant.  This factorization property was first established via a different method (that we will compare against in Sec.~\ref{sec:Comparison} below) in Ref.~\cite{Lupsasca2018}.  As noted therein, the throat has a logarithmically divergent radial proper depth $\sim\ab{\log{\kappa}}$ in the limit $\kappa\to0$ of extremal spin (Fig.~\ref{fig:NHEK} bottom right).  Hence, a disk with a uniform density of emitters per unit proper radial distance could contribute a flux
\begin{align}
    F_\mathrm{disk}^\mathrm{NHEK}(g)\stackrel{\kappa\to0}{\sim}F_g\ab{\log{\kappa}}
\end{align}
from its portion in the throat (but still outside the ISCO).  This means the critical flux curve \eqref{eq:CriticalFlux} may be observable in principle, though likely not in practice as this would require collecting photons over the entire celestial sphere (or from many randomly oriented high-spin BHs).

\section{Horizon and celestial sphere unfolded in the emitter sky}
\label{sec:Unfolding}

\begin{figure*}
	\centering
	\vspace{-10pt}
	\begin{overpic}[percent,width=\textwidth]{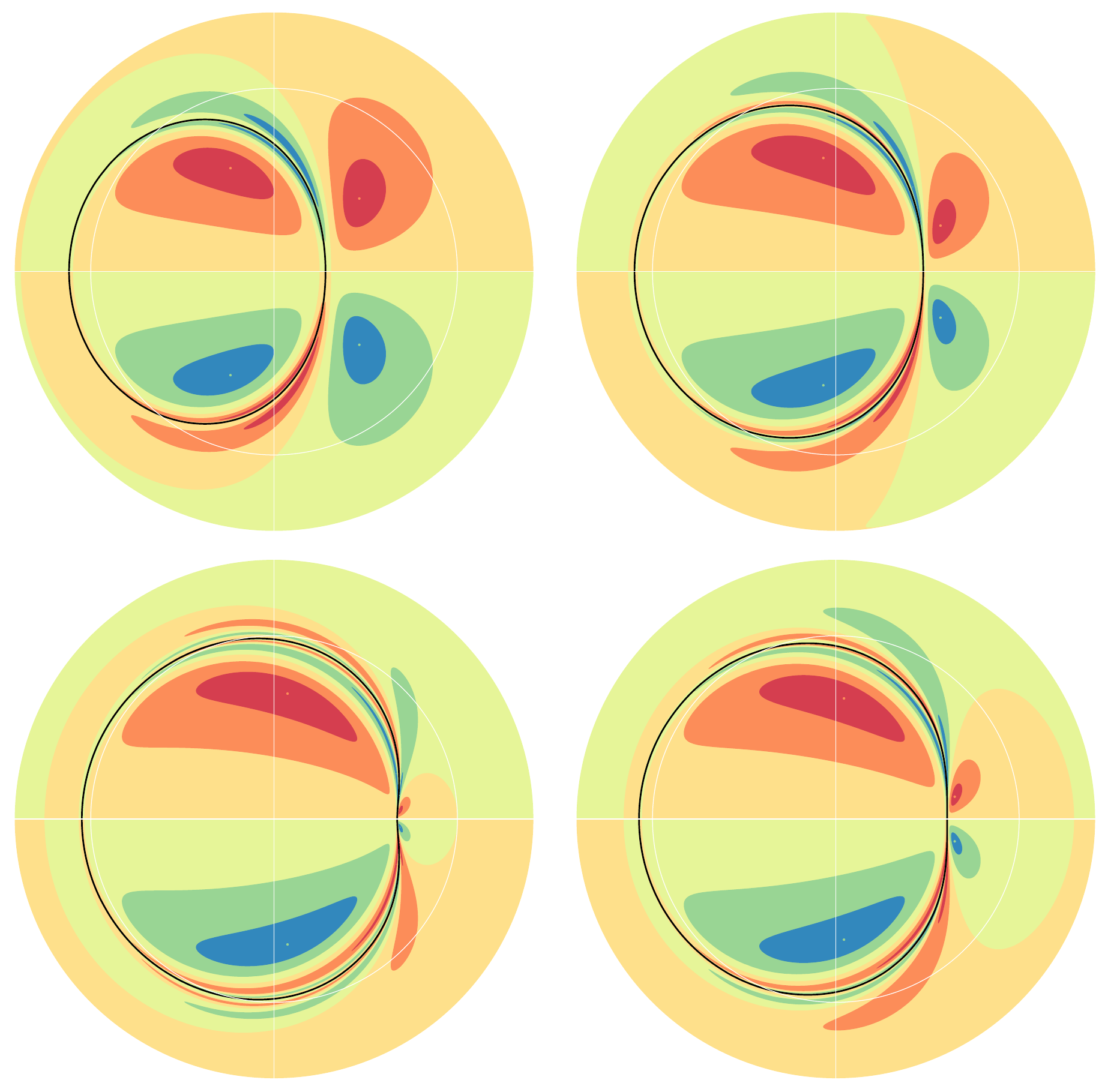}
		\put(42,42){\includegraphics[width=.16\textwidth]{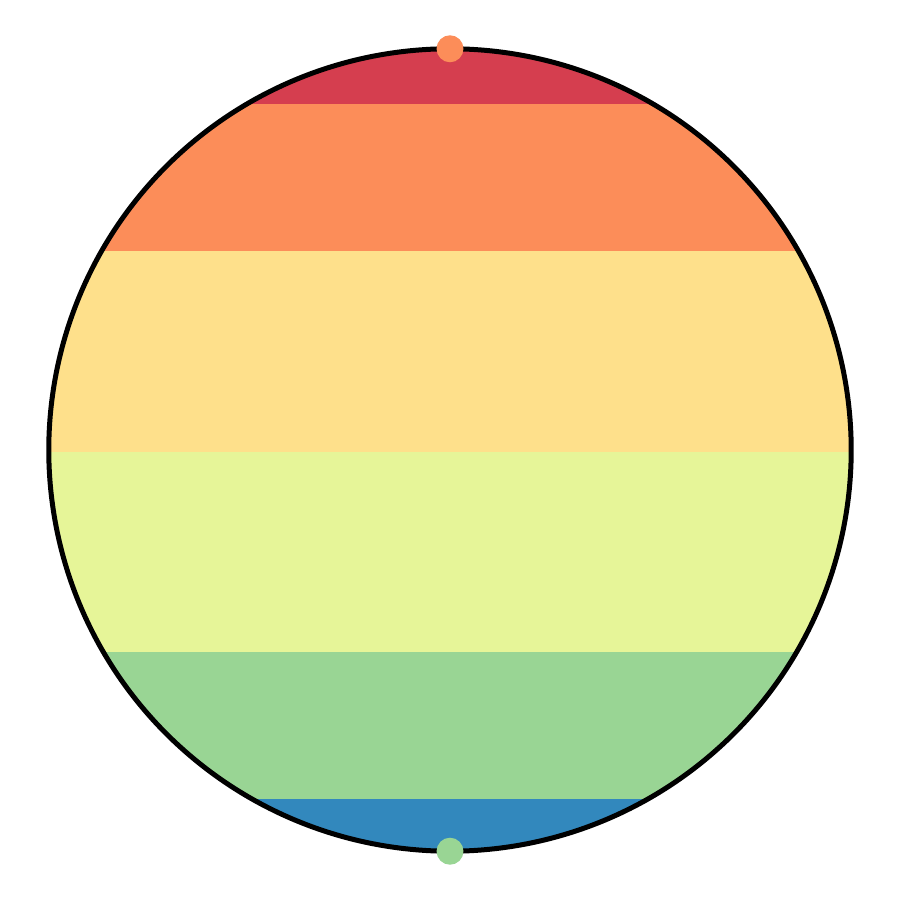}}
	\end{overpic}
	\caption{Latitude bands of the event horizon and celestial sphere, as seen in the sky of an equatorial orbiter on the ISCO.  Rays emitted inside the critical curve (black) eventually cross the horizon, while rays emitted outside eventually reach the celestial sphere.  We show the position of these rays in the area-preserving backside projection \eqref{eq:BacksideProjection} of the emitter sky [Fig.~\ref{fig:BacksideProjection}], colored by the latitude of emission on the event horizon or celestial sphere, as shown in central inset (colors change every $30^{\circ}$, with orange/green dots depicting the north/south poles).  The orbiter sees infinitely many ``unfoldings'' of both the horizon and the celestial sphere.  Here, we show the sky of a prograde ISCO orbiter ($r_s=r_\mathrm{ms}^+$) for black hole spins (clockwise from top left) $a/M=90\%$, $a/M=98\%$, $a/M=99.8\%$, and $a/M=99.99\%$.  As the spin approaches extremality, the unfoldings shift away from the critical curve, turning like pages in a book.  Analogous plots for a distant observer are displayed in Fig.~2 of Ref.~\cite{GrallaLupsasca2020a}.}
	\label{fig:Unfolding}
\end{figure*}

In the previous sections, we have considered isotropic emitters on circular equatorial orbits around a Kerr BH.  We now ask: where does their emission go?  To answer this question, one must understand the behavior of rays emitted from an orbiter.  We begin by studying the eventual fate of these light rays as a function of their emission direction, and find that the horizon and celestial sphere are ``unfolded'' many times in the sky (Fig.~\ref{fig:Unfolding}).  In Sec.~\ref{sec:FluxCurves}, we computed the total flux radiated by an emitter to the celestial sphere; in this section, we also discuss how this calculation may be extended to a computation of the flux density $F_g(\theta_o)$ received at infinity as a function of the observer inclination $\theta_o$.  In general, this is a hard problem requiring a numerical study of lensing by the BH.  However, Ref.~\cite{Lupsasca2018} noted that certain simplifications occur for the special case of emission from the near-horizon region of a high-spin BH, and obtained an approximate analytic formula for $F_g(\theta_o)$ in that regime.  Here, we revisit this result in light of our exact analytic formula \eqref{eq:CriticalFlux}, allowing us to clarify the status of its underlying approximation.

\subsection{Behavior of rays emitted from an orbiter}

In this section, we study the eventual fate of photons emitted from a circular equatorial Kerr orbiter as a function of emission direction $(\Theta,\Psi)$ in the local sky.  In particular, we consider an orbiter on the prograde ISCO ($r_s=r_\mathrm{ms}^+$) and wish to determine the angle $\theta_o(\Theta,\Psi)$ at which a given light ray either reaches the celestial sphere $r_o\to\infty$ (if it escapes) or else crosses the event horizon $r_+=M+\sqrt{M^2-a^2}$ (if it is captured by the BH).  Since the spacetime is stationary and axisymmetric, this calculation only requires the poloidal $(r,\theta)$ component of the Kerr null geodesic equation,
\begin{align}
	\label{eq:NullGeodesics}
	I_r=\fint_{r_s}^{r_o}\frac{\ed r}{\pm_r\sqrt{\mathcal{R}(r)}}
	=\fint_{\theta_s}^{\theta_o}\frac{\ed\theta}{\pm_\theta\sqrt{\Theta(\theta)}}
	=G_\theta,
\end{align}
where $(r_s,\theta_s)=(r_\mathrm{ms}^+,\pi/2)$ is the emitter position, while $\mathcal{R}(r)$ and $\Theta(\theta)$ are the radial and angular potentials \eqref{eq:RadialPotential}--\eqref{eq:AngularPotential} expressed in terms of the energy-rescaled conserved quantities $(\lambda,\eta)$ [Eq.~\eqref{eq:EnergyRescaledQuantities}].  The signs $\pm_{r,\theta}$ correspond to the direction of radial and polar motion, and the slashes across the integral signs indicate that these signs switch at every radial and polar turning point.  

The analytic solution of this equation was recently given in a convenient form involving elliptic integrals in Ref.~\cite{GrallaLupsasca2020b}, allowing for simple studies of the optical appearance of Kerr equatorial disks as viewed by a distant observer at infinity \cite{GrallaLupsasca2020a,GrallaLupsasca2020d,Gates2020}.  For such an observer, it has become standard to use Cartesian coordinates on the sky $(\alpha,\beta)$ that were first introduced by Bardeen \cite{Bardeen1973}, and illustrations of the behavior of light rays as a function of these impact parameters were given in Fig.~2 of Ref.~\cite{GrallaLupsasca2020a}.  

Since the emitter sky is really a sphere, no such coordinate system is available to us here, so we use the backside projection \eqref{eq:BacksideProjection} instead. (We do not use the redshift-weighted projection \eqref{eq:FrontsideProjection}, as it excludes part of the photon capture region in the sky.)  That is, each point $(\rho,\varphi)$ on the plane corresponds via Eq.~\eqref{eq:BacksideProjection} to an emission angle $(\Psi,\Upsilon)$, which in turn corresponds to a choice of $(\lambda,\eta)$ [Eqs.~\eqref{eq:PolarAngle}--\eqref{eq:Upsilon}].  Specializing to the present case of equatorial sources, we can exclude vortical geodesics (with $\eta<0$) that can never reach the equator \cite{Kapec2020,GrallaLupsasca2020b}.  In that case, inverting Eq.~\eqref{eq:NullGeodesics} results in
\begin{align}
    \label{eq:PolarTrajectory}
    \cos{\theta_o}=\mp_\theta\sqrt{u_+}\sn\pa{\left.\sqrt{-u_-a^2}I_r\right|\frac{u_+}{u_-}},
\end{align}
where $\pm_\theta=\sign\pa{\sin{\varphi}}=+/-$ according to whether the photon is emitted in the lower/upper half-plane of the projection, $\sn(z|k)$ is the Jacobi elliptic sine function, and $u_\pm$ denote the roots of the angular potential,
\begin{align}
	u_\pm=\triangle_\theta\pm\sqrt{\triangle_\theta^2+\frac{\eta}{a^2}},\quad
	\triangle_\theta=\frac{1}{2}\pa{1-\frac{\eta+\lambda^2}{a^2}}.
\end{align}
The last ingredient needed to compute $\theta_o(\rho,\varphi)$ using Eq.~\eqref{eq:PolarTrajectory} is the radial integral $I_r$ evaluated along the full extent of the radial trajectory.  Photons emitted inside/outside the critical curve eventually reach the horizon/infinity, either directly (if $\pm_r=-/+$) or after encountering a radial turning point (if $\pm_r=+/-$).  The contribution to $I_r$ from each segment of the radial motion is analytically given in App.~A of Ref.~\cite{GrallaLupsasca2020a}.

Putting everything together, we plot in Fig.~\ref{fig:Unfolding} contours of constant $\theta_o(\rho,\varphi)$ in the backside-projected sky of an ISCO emitter orbiting BHs of increasingly high spins.  An observer on such an orbit would manifestly perceive a highly distorted view of the surrounding spacetime due to the strong lensing by the nearby BH.  We now highlight some qualitative features of this lensing.

Critical light rays with $(\lambda,\eta)=(\tilde{\lambda},\tilde{\eta})$ asymptote to (unstable) bound orbits in the photon shell.  Near-critical light rays can orbit multiple times around the BH before reaching the celestial sphere or horizon, with the number $m$ of librations (polar oscillations) along the path diverging logarithmically in the deviation from criticality \cite{Johnson2019,GrallaLupsasca2020a}.  As a result, the orbiter sees infinitely many images of both the horizon and the celestial sphere ``unfolded'' in the sky, with each image labeled by an integer $m$.  Since the equatorial orbiter preserves the equatorial reflection symmetry of the underlying spacetime, the contours in Fig.~\ref{fig:Unfolding} are symmetric about the horizontal line marking the equatorial plane. 

The direct $m=0$ image is usually the most prominent in the sky, with more strongly lensed images appearing as exponentially demagnified lobes, squeezed exponentially closer to the critical curve.  In the NHEK regime, as the orbiter is pushed deeper into the throat of a high-spin BH, higher-$m$ lobes become increasingly prominent, with each unfolding of the celestial sphere turning from left to right like pages in a book (see also next section).

Note that outside the NHEK regime, there are photons that connect to infinity, traveling along paths with one radial turning point but no polar turning points.  This is evident in the top left panel of Fig.~\ref{fig:Unfolding}, where the $m=0$ image of the celestial sphere reaches into the left half of the projected sky (corresponding to photons initially shot radially inwards by the emitter).

\subsection{Computing observed flux in the emitter frame}

The study of thin equatorial disks in General Relativity (GR) has a long history stretching back to the seminal works of Novikov and Thorne \cite{Novikov1973} and Page and Thorne \cite{Page1974}.  Soon after these papers, Cunningham \cite{Cunningham1975} initiated an analysis of the effects of GR on the propagation of radiation from the disk to a distant observer, and Luminet \cite{Luminet1979} produced a simulated image of such a disk.

Typically, the properties of radiation received from the disk at infinity are analyzed in the frame of a distant observer, considering only the light rays that connect observer and disk.  In this usual approach, therefore, one shoots light rays back from a fixed observer to hit every possible point on the disk.  However, it is in principle possible to do the computation in reverse, shooting light rays in every direction from each orbiter in the disk and keeping track of those that reach the observer at infinity.  In practice, this approach is highly inefficient and technically challenging, especially if one wishes to study time-dependent effects.  Nonetheless, if one is only interested in addressing time-independent questions or computing time-averaged quantities (averaged over the period of an emitter's orbit), then a graphical method based on the projections that we have defined becomes tractable.

For instance, several properties of the emission can be determined graphically using contour plots of $\cos{\theta_o}$ in the backside projection, such as those depicted in Fig.~\ref{fig:Unfolding}.  As an example, one can compute $g_\mathrm{max}(r_s,m)$, the maximum blueshift of photons received by a far observer at inclination $\theta_o$ after encountering $m$ polar turning points on their way from an equatorial emitter at source radius $r_s$.  First, one graphically finds the radius $\rho$ of the largest circle in the projection that touches the contour of $\theta_o$ in the $m^\text{th}$ image of the celestial sphere.  Then, one determines the corresponding emission angle $\Psi(\rho)$ using the inverse of Eq.~\eqref{eq:BacksideProjection}, before plugging it into the formula \eqref{eq:RedshiftPsi} for $g(\Psi)$.  The resulting quantity $g_\mathrm{max}(r_s,m)$ was recently investigated in Ref.~\cite{Gates2020} from the opposite perspective, by considering images of the disk labeled by $m$ as seen by the observer (see Figs.~3 and 4 therein).  Both approaches are comparable in difficulty, and render different features manifest.  For instance, in the emitter frame, it is clear that the maximally blueshifted photons are always emitted in the direction of orbital motion, and therefore received only by equatorial observers.  It is also easy to read off on which image $m$ the equatorial observer will see these bluest photons.  In the NHEK regime \eqref{eq:DoubleLimit}, this label is analytically given by $m\approx\lfloor m_0(\kappa)\rfloor$, where\footnote{From Ref.~\cite{Gralla2017}, we take Eq.~(4.3) and set $\hat{G}_\theta=0$ for an equatorial observer (App.~D), $D_o\sim R_o\to\infty$ for a far observer [Eq.~(3.20)], and $R_s=\bar{R}$ with $\epsilon=\kappa^p$ [Eq.~(3.2)].  The Carter constant is $q=\sqrt{3}$ for equatorial rays [Eq.~(3.13)], and hence $qG_\theta=\pi$ (App.~D).}
\begin{align}
    \label{eq:PageNumber}
	m_0=\frac{1}{\pi}\log\pa{\frac{36}{\pa{2+\sqrt{3}}R_s}}-\frac{p}{\pi}\log{\kappa}.
\end{align}
This was numerically checked to correspond to the emission with maximum blueshift \cite{Gates2020} and maximum flux \cite{Gralla2017}.

As another example, one can also compute the total flux $F_g(\theta_o)$ received with redshift $g$ at observation inclination $\theta_o$ via graphical methods.  To do so, one plots on the projected emitter sky circles of constant $\hat{g}\lessgtr g^{(i)}\lessgtr g_0$ as well as contours of constant $0<\theta_o^{(j)}<\pi$ in the $m^\text{th}$ image of the celestial sphere [Eq.~\eqref{eq:PolarTrajectory}], resulting in a tiling of the projected sky.  Setting $F_N=4\pi$, the flux is
\begin{align}
    \label{eq:FluxSum}
    F_{\hat{g}^{(i)}}(\hat{\theta}_o^{(j)})&=\sum_{m=0}^\infty F_{\hat{g}^{(i)}}^{(m)}(\hat{\theta}_o^{(j)}),\\
    F_{\hat{g}^{(i)}}^{(m)}(\hat{\theta}_o^{(j)})&\approx\hat{g}^{(i)}\mathcal{A}^{(m)}(\hat{g}^{(i)},\hat{\theta}_o^{(j)})
    \approx\mathcal{A}_g^{(m)}(\hat{g}^{(i)},\hat{\theta}_o^{(j)}),\notag
\end{align}
where $\hat{x}^{(i)}\equiv(x^{(i+1)}-x^{(i)})/2$, while $\mathcal{A}^{(m)}(\hat{g}^{(i)},\hat{\theta}_o^{(j)})$ and $\mathcal{A}_g^{(m)}(\hat{g}^{(i)},\hat{\theta}_o^{(j)})$ denote the areas on the backside and redshift-weighted projections \eqref{eq:BacksideProjection} and \eqref{eq:FrontsideProjection}, respectively, of the cell bounded by the contours of $g^{(i)}$, $g^{(i+1)}$, $\theta_o^{(j)}$, and $\theta_o^{(j+1)}$ in the $m^\text{th}$ image lobe.  This approximation improves as $g^{(i)}$ and $\theta_o^{(j)}$ are sampled more densely, resulting in a finer mesh of tiles, and becomes exact in the limit where $g^{(i)}$ and $\theta_o^{(j)}$ are continuous and the grid cells become infinitely small.

In practice, the $m\ge2$ contributions to the sum \eqref{eq:FluxSum} are vanishingly small outside the NHEK regime \eqref{eq:DoubleLimit}, and the flux is dominated by the direct $m=0$ rays, with sometimes significant corrections from $m=1$ rays.  This type of graphical calculation was carried out for extreme Kerr emitters by Cunningham and Bardeen \cite{Cunningham1973}, with results displayed in their Fig.~5.\footnote{Their flux seems to decrease as $r_s\to M$, presumably because they only focused on the $m=0$ and $m=1$ contributions, whereas higher lobes become dominant in this regime.  This may explain the apparent contradiction we noted above between our analytic result \eqref{eq:CriticalFlux} and their vanishing flux as $r_s\to M$ in their Fig.~6.}  They used the slightly more convenient redshift-weighted projection, i.e., they counted $\mathcal{A}_g^{(m)}$.  In the NHEK regime, however, the dominant lobe in the emitter sky has increasingly large $m$.  Hence, the sum \eqref{eq:FluxSum} can no longer be truncated at small $m$, and a different approach is required (see next section).

Finally, for time-averaged observables that do not depend on angle of escape at infinity, such as the total escape probability (Sec.~\ref{sec:EscapeProbability}) and total integrated flux collected on the celestial sphere (Sec.~\ref{sec:FluxCurves}), the emitter-frame approach is dramatically simpler than the observer-frame approach: the latter would require computing observer-dependent observables at every inclination $\theta_o$ before integrating them over the celestial sphere, while the former is analytically tractable and therefore highly preferable.

\subsection{Angle-dependent flux from a NHEK emitter}

In Sec.~\ref{sec:FluxCurves}, we derived an exact formula \eqref{eq:CriticalFlux} for the total flux radiated to the entire celestial sphere by an individual NHEK emitter.  Since this result was independent of the orbital radius $R_s$ in NHEK, we concluded that the observable $F_g$ displays critical behavior in NHEK, where it reaches a nonvanishing fixed point.  Ideally, we would like to use this emitter-frame approach to graphically compute the more fine-grained, angle-dependent quantity $F_g(\theta_o)$ for a NHEK emitter using Eq.~\eqref{eq:FluxSum}.  Unfortunately, this turns out to be very difficult; in fact, $F_g(\theta_o)$ does not even converge to a fixed point in the NHEK regime, as we now explain.

Different regions of the emitter sky are labeled by the number $m$ of polar turning points encountered by the emanating rays.  As $\kappa\to0$ in the regime \eqref{eq:DoubleLimit} of increasing BH spin, the emitter is pushed deeper into the throat and the $m=0$ region shrinks, while the $m=1$ image begins to dominate the sky.  Increasing the spin further, the $m=1$ image itself begins to shrink and the $m=2$ image takes over.  This transition process repeats ad infinitum, with the label $m\approx\lfloor m_0\rfloor$ of the dominant lobe ever increasing as we dial the BH up to extremality [Eq.~\eqref{eq:PageNumber}].  This process is illustrated in Fig.~\ref{fig:Unfolding}: as $\kappa\to0$ (clockwise direction from top left), $\theta_o$ contours turn like ``pages in a book'', flowing from left (where they start out bunched exponentially close to the critical curve) to right.  The ``open page'' $m\approx\lfloor m_0\rfloor$ accounts for the majority of the flux, with contributions from other pages exponentially suppressed (Fig.~4 of Ref.~\cite{Gralla2017}).\footnote{This behavior holds more generally: the flux from each successive lensed image is demagnified by $e^{-\gamma}$, with $\gamma$ a Lyapunov exponent characterizing the orbital instability of critical photons \cite{Gralla2019,Johnson2019,GrallaLupsasca2020a}.}  Thus, while there is a critical fixed point \eqref{eq:CriticalFlux} for the flux radiated to infinity,
\begin{align}
    \label{eq:ObserverIntegral}
    F_g=\int_0^\pi F_g(\theta_o)\sin{\theta_o}\ed\theta_o,
\end{align}
as required by conservation of energy, there is no limiting fixed point for the flux $F_g(\theta_o)$ received at each angle on the celestial sphere.  Strikingly, while varying the NHEK radius of the emitter can significantly alter its flux to infinity $F_g(\theta_o)$, this variation must nonetheless always conspire to preserve the integral \eqref{eq:ObserverIntegral}.

A similar effect arises in computations of gravitational-wave emission from near-horizon processes \cite{Porfyriadis2014,Hadar2014,Gralla2015} where, fixing the retarded time, the overall phase of the wave oscillates and does not settle down as the extremal limit $\kappa\to0$ is taken.  Intuitively, this is caused by the ``stretching of the NHEK'', and the energy flux is unaffected by this overall oscillating phase.

Together with the page-turning analogy and the conservation of the integral \eqref{eq:ObserverIntegral}, this observation suggests that $F_g(\theta_o)$ varies quasiperiodically with the NHEK radius $R_s$.  Next, we will integrate $F_g(\theta_o)$ over a range of nearby radii $R_s$ to obtain an approximate formula [Eq.~\eqref{eq:ApproximateCriticalFluxDensity} below] for the flux from a thin disk in NHEK.

\subsection{Angle-dependent flux from a NHEK disk}
\label{sec:Comparison}

In this section, we discuss $F_\mathrm{disk}^\mathrm{NHEK}(g,\theta_o)$: the flux density received at observer inclination $\theta_o$ and redshift $g$ from the NHEK portion of a disk.  This quantity was derived in Ref.~\cite{Lupsasca2018} using a semianalytic approximation justified by the results of Ref.~\cite{Gralla2017}.  We now briefly summarize the derivation, before testing the validity of its underlying approximation by integrating $F_\mathrm{disk}^\mathrm{NHEK}(g,\theta_o)$ over the celestial sphere and comparing against our exact result from Sec.~\ref{sec:FluxCurves}.

Since the quantity $F_\mathrm{disk}^\mathrm{NHEK}(g,\theta_o)$ involves different emission radii, it is more natural to compute it in the observer frame, by examining images of the equatorial disk labeled by $\bar{m}$ (see Ref.~\cite{Gates2020} for a recent discussion).  In this approach, one sums over images of the disk,
\begin{align}
    \label{eq:ObserverFluxSum}
    F_\mathrm{disk}(g,\theta_o)=\sum_{\bar{m}=0}^\infty F_\mathrm{disk}^{(\bar{m})}(g,\theta_o).
\end{align}
In practice, it is always necessary to truncate this sum.  Outside the throat, there is negligible emission at large $\bar{m}$, and we only need to take into account the $\bar{m}=0$ image (and possibly $\bar{m}=1$) containing most of the flux,
\begin{align}
    \label{eq:FirstImage}
    F_\mathrm{disk}(g,\theta_o)\approx F_\mathrm{disk}^{(0)}(g,\theta_o).
\end{align}
Inside the throat, on the other hand, the dominant contribution comes from $\bar{m}\approx\lfloor m_0\rfloor$, where $m_0\sim\log\kappa^{-p}$ is given in Eq.~\eqref{eq:PageNumber} and scales with the NHEK radius.  The flux is then better approximated by the truncation
\begin{align}
    \label{eq:DiagonalCut}
    F_\mathrm{disk}^\mathrm{NHEK}(g,\theta_o)
    &\approx F_\mathrm{disk}^{(m_0)}(g,\theta_o),
\end{align}
instead of Eq.~\eqref{eq:FirstImage}.  This takes a ``diagonal cut'' across the sum \eqref{eq:ObserverFluxSum} in the sense that, as one goes to deeper radii, one also increases $m_0$, thereby considering higher images.\footnote{This effectively treats $\bar{m}$ as a continuous parameter.  Since image labels must be integers, this approximation can only be ``exact'' when $m_0$ takes integer values.  The intersections of the orange and blue curves in Fig.~\ref{fig:DiskEnvelope} occur precisely whenever that happens.}  Under this approximation, Ref.~\cite{Lupsasca2018} found that
\begin{align}
    \label{eq:ApproximateCriticalFluxDensity}
    F_\mathrm{disk}^\mathrm{NHEK}(g,\theta_o)&\stackrel{\kappa\to0}{\approx}\frac{\mathcal{N}}{\sin{\theta_o}}\frac{\pa{g-\frac{1}{\sqrt{3}}}\pa{g+\sqrt{3}}}{\sqrt{\mathcal{Q}+\cos^2{\theta_o}-4\cot^2{\theta_o}}},\notag\\
    \mathcal{Q}&=3-\frac{3}{4g^2}\pa{g-\frac{1}{\sqrt{3}}}\pa{5g+\sqrt{3}},
\end{align}
where the proportionality factor $\mathcal{N}$ only depends on an integral over NHEK radii and will be fixed momentarily.

We may now integrate this quantity over the celestial sphere, as in Eq.~\eqref{eq:ObserverIntegral}.  This results in a complete elliptic integral over observer inclinations, of the form computed in Eq.~(80a) of Ref.~\cite{Kapec2020} with $Q=\mathcal{Q}$, $P=1$ and $\ell^2=4$:
\begin{align}
    \label{eq:ApproximateCriticalFlux}
    F_\mathrm{disk}^\mathrm{NHEK}(g)&\approx\frac{\mathcal{N}\pa{g-\frac{1}{\sqrt{3}}}\pa{g+\sqrt{3}}}{\sqrt{-U_-}}K\pa{\frac{U_+}{U_-}},\\
    U_\pm&=\frac{\pm\sqrt{(\mathcal{Q}+1)(\mathcal{Q}+9)}-(\mathcal{Q}+3)}{2}.
\end{align}
This approximation, which has support $1/\sqrt{3}\le g\le\sqrt{3}$, can almost be compared against our exact result for the total flux received from the NHEK portion of a disk, given in Eq.~\eqref{eq:FluxDiskNHEK}, or more simply, Eq.~\eqref{eq:CriticalFlux} with $F_N$ taken to be the Newtonian disk flux.  The only missing ingredient is the value of $\mathcal{N}$.  We conjecture that the correct choice to carry out the comparison is\footnote{Note that Eq.~(3.10) of Ref.~\cite{Lupsasca2018} gives $\mathcal{N}=\frac{3\sqrt{3}}{8r_o^2}\int_\mathrm{NHEK}\mathcal{E}(r_s)\ed r_s$.} 
\begin{align}
    \label{eq:}
    \mathcal{N}=\frac{3\sqrt{3}}{8\pi}F_N.
\end{align}
This choice ensures that the exact and approximate formulas for $F_\mathrm{disk}^\mathrm{NHEK}(g)$ precisely match at $g=\sqrt{3}$.  The next section is entirely devoted to an argument in support of this matching condition.  Before turning to it, we note that Eqs.~\eqref{eq:CriticalFlux} and \eqref{eq:ApproximateCriticalFlux} are compared in Fig.~\ref{fig:CriticalFluxCurve}.  Integrating over redshift $g$, we find that the approximate formula \eqref{eq:ApproximateCriticalFlux} for the flux accounts for about $60.1\%$ of the exact flux \eqref{eq:CriticalFlux}.  As expected, our truncation missed some contributions to the flux, which we now describe.

\subsection{Light rays connecting NHEK to infinity}

In this section, we further explore the nature of the flux contributions taken into account by the approximation \eqref{eq:ApproximateCriticalFluxDensity}.  In the NHEK regime, there are two different types of rays connecting NHEK emitters to the celestial sphere:
\begin{itemize}
    \item Type I: a measure zero set of rays, each providing a very large contribution to the flux, corresponding to directions of emission such that the orbital radius is a turning point of the radial trajectory \cite{Gralla2017,Lupsasca2018}.
    \item Type II: the other rays, a finite-measure set, each providing a vanishingly small flux contribution.
\end{itemize}
Integrating over the infinite depth of the throat results in finite flux from both Type I and Type II contributions.  We will now argue that the approximation \eqref{eq:ApproximateCriticalFluxDensity} gives precisely the flux received from the Type I rays.

\begin{figure}
	\centering
	\includegraphics[width=\columnwidth]{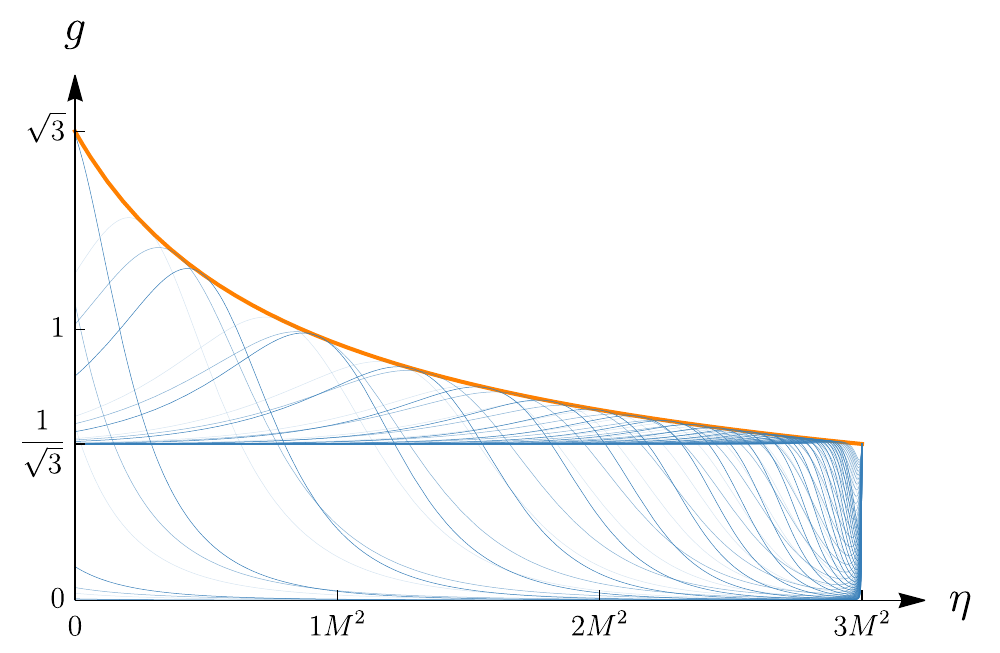}
 	\caption{Observed redshift $g(\eta,m,R_s)$ of photons emitted from NHEK orbiters around a near-extremal BH with spin $a=M\sqrt{1-10^{-18}}$.  The blue curves correspond to light rays that connect a source ring $R_s$ in the NHEK to an equatorial observer at infinity with $m\in\br{0,30}$ polar turning points (see Ref.~\cite{Gates2020} for details).  We show $R_s=R_\mathrm{ms}=2^{1/3}$ (solid blue) together with $R_s=R_\mathrm{ms}+5$ and $R_s=R_\mathrm{ms}+10$ (transparent blue), and overlay the envelope $g_0(\eta)$ [Eq.~\eqref{eq:RedshiftEnvelope}] accounting for the light rays with dominant flux (orange).}
	\label{fig:DiskEnvelope}
\end{figure}

It is instructive to plot $g(\eta,m,R_s)$, the observed redshift of light rays connecting an emitter at $R_s$ to infinity (after $m$ angular turning points) as a function of their energy-rescaled Carter constant $\eta$ (blue curves in Fig.~\ref{fig:DiskEnvelope}).  We can think of $\eta$ as a parameter along the NHEKline, as seen by the far observer \cite{Lupsasca2018,Gates2020}.  These curves depend on the observer inclination.  In Fig.~\ref{fig:DiskEnvelope}, for example, we chose an equatorial observer $\theta_o=\pi/2$.

The redshift  has a complicated quasiperiodic dependence on $R_s$: as the emission radius varies, the blue curves (of fixed $m$) sweep the area under their orange envelope, and some shifts in $R_s$ can almost map a curve of fixed $m$ into another curve of fixed $m+1$.

The orange envelope is obtained by plugging $m=m_0$ into $g(\eta,m,R_s)$, which eliminates its dependence on the NHEK radius $R_s$ and results in (see Ref.~\cite{Gates2020} for details)
\begin{align}
	\label{eq:RedshiftEnvelope}
	g_0=\frac{\sqrt{3}}{2\sqrt{1+\eta/M^2}-1}.
\end{align}
The Type I rays correspond precisely to the intersections of the blue curves with their orange envelope, indicating that a photon is emitted from its radial turning point (see Eq.~(4.3) of \cite{Gralla2017}).  Note that every point in the plane $(\eta,g)$ under the orange envelope corresponds to an infinite number of Type II light rays connecting some discrete set of NHEK emitters to a far observer.

We can use Fig.~\ref{fig:DiskEnvelope} to graphically explain some features of the emission from a NHEK disk.  First, note that, for fixed $g$ (corresponding to a horizontal cut in Fig.~\ref{fig:DiskEnvelope}), there are generically multiple rays that connect emitter and observer (corresponding to points where our horizontal cut intersects blue curves).  We can separately count the numbers $n_\mathrm{I}(g,R_s)$ and $n_\mathrm{II}(g,R_s)$ of (special) Type I and (generic) Type II rays, respectively, that are emitted from radius $R_s$ and reach the observer with redshift $g$:
\begin{itemize}
    \item
    For $g\in(0,1/\sqrt{3})$, $n_\mathrm{I}(g,R_s)=0$ and $n_\mathrm{II}(g,R_s) > 0$.
    \item
    At the crossing of $g=1/\sqrt{3}$, $n_\mathrm{I}(g,R_s)=0$, while  
    \begin{align}
       \frac{{\lim\limits_{g\to\pa{1/\sqrt{3}}^+}}n_\mathrm{II}(g,R_s)}{\lim\limits_{g\to\pa{1/\sqrt{3}}^-}n_\mathrm{II}(g,R_s)}&\stackrel{\kappa\to0}{\approx}2.
    \end{align}
    This is precisely the origin of the NHEKline in $F_g$ at criticality [Eq.~\eqref{eq:CriticalFlux}].  Across the discontinuity at $g=1/\sqrt{3}$, $F_g$ must precisely increase by a factor of 2, as seen in Eq.~\eqref{eq:Discontinuity}.  
    \item
    For $g\in(1/\sqrt{3},\sqrt{3})$, $n_\mathrm{I}(g,R_s)=1$ if $m_0$ is an integer (and zero otherwise), whereas $n_\mathrm{II}(g,R_s)\geq0$.
    \item At $g=\sqrt{3}$, $n_\mathrm{I}(g,R_s)=1$ if $m_0$ is an integer (and zero otherwise) and $n_\mathrm{II}(g,R_s)=0$.
\end{itemize}

When we integrate over many NHEK radii (i.e., a disk of NHEK emitters), we fill all points on and under the orange envelope \eqref{eq:RedshiftEnvelope} with rays that connect the disk to the observer.  In particular, since the approximation \eqref{eq:DiagonalCut} is by definition exact for Type I rays, it follows that the flux contribution from Type I is given by Eq.~\eqref{eq:ApproximateCriticalFluxDensity}.

For NHEK emitters, the photons with the maximum blueshift $g=\sqrt{3}$ come only from rays in the forward direction, which are always of Type I and thus included in the Type I flux.  We therefore ought to set the total spectral flux \eqref{eq:CriticalFlux} and the Type I spectral flux \eqref{eq:ApproximateCriticalFlux} to be equal at that redshift: $F_g=F_\mathrm{disk}^\mathrm{NHEK}(g)$ at $g=\sqrt{3}$.  This is the origin of the matching condition in the previous section, where we found that the flux from the Type I rays accounts for $\approx60.1\%$ of the total flux.  While there is no analytic method for calculating the Type II flux independently, directly reproducing its total contribution stands a litmus test for further methods of calculation.

\section{High-spin perturbation theory}
\label{sec:PerturbationTheory}

\begin{table}
\renewcommand{\arraystretch}{2}
\begin{tabular}{||c||c|c|c|c|c|c||}
\hline
\multicolumn{7}{|c|}{Extremal value \& leading correction as $\kappa\to0$} \\
\hline\hline
NHEK $p$-band  & \textcolor{green}{$(0,\frac{q}{2})$} & \textcolor{blue}{$\frac{q}{2}$} & \textcolor{purple}{$\pa{\frac{q}{2},q}$} & \textcolor{purple}{$q$} & \textcolor{purple}{$(q,1)$} & \textcolor{red}{$1$} \\
\hline
Extreme $\cos{\tilde{\Theta}}$ & $0$ & $\kappa^0$ & \multicolumn{3}{c|}{$\kappa^0$} & $\kappa^0$ \\
\hline
Correction & $\kappa^{-2p+q}$ & $\kappa^{q/2}$ & $\kappa^{2p-q}$ & $\kappa^q$ & $\kappa^q,\kappa^{2(1-p)}$ & $\kappa^q,\kappa^{1-q}$ \\
\hline
Extreme $\cos{\tilde{\Psi}}$ & $\kappa^0$ & $\kappa^0$ & \multicolumn{3}{c|}{$0$} & $0$\\
\hline
Correction & $\kappa^{-2p+q}$ & $\kappa^q$ & $\kappa^{2p-q}$ & $\kappa^q$ & $\kappa^q,\kappa^{2-p-q}$ & $\kappa^q,\kappa^{1-q}$ \\
\hline
\end{tabular}
\caption{The leading corrections to the direction cosines $\cos{\tilde{\Theta}}$ and $\cos{\tilde{\Psi}}$ in the NHEK regime \eqref{eq:DoubleLimit} obey different scaling behaviors in $\kappa$ depending on the values of $p$ and $q$.  To obtain a good approximation to the critical curve at subleading order (Fig.~\ref{fig:SubleadingCriticalCurve} top), we glue segments corresponding to NHEK bands where the exponents of $\kappa$ change their functional form, breaking the power law scalings (Fig.~\ref{fig:SubleadingCriticalCurve} bottom).}
\label{tbl:Scalings}
\end{table}

\begin{figure}
	\centering
	\includegraphics[width=\columnwidth]{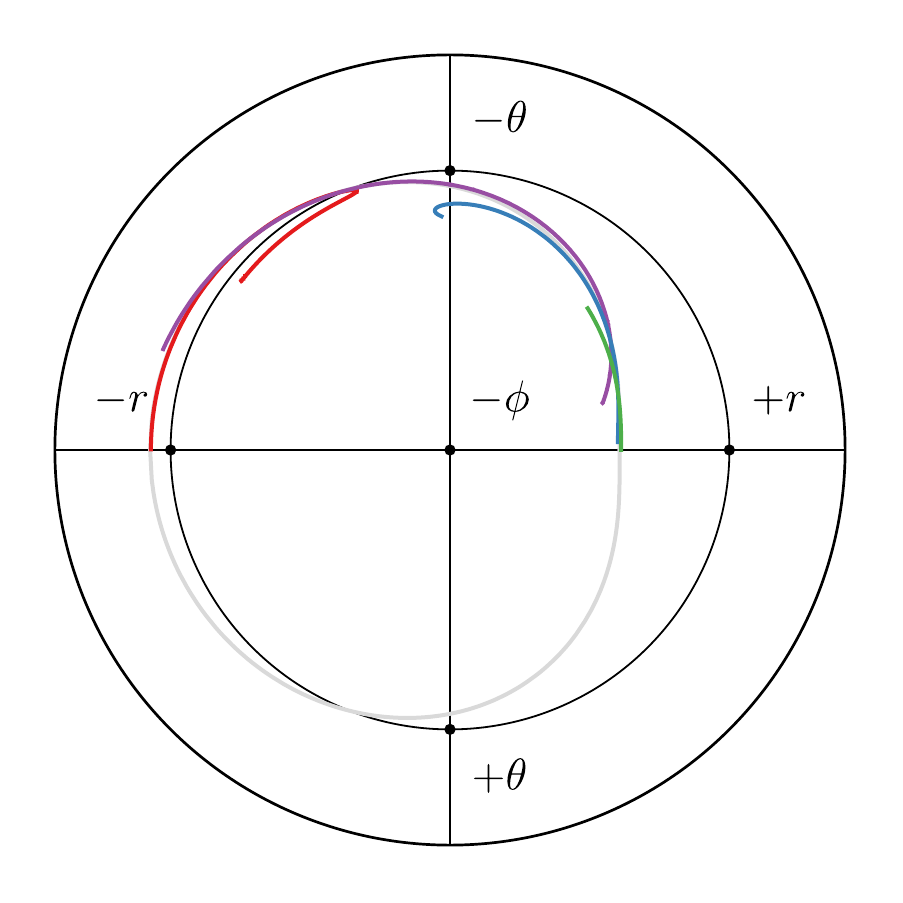}\\
	\includegraphics[width=\columnwidth]{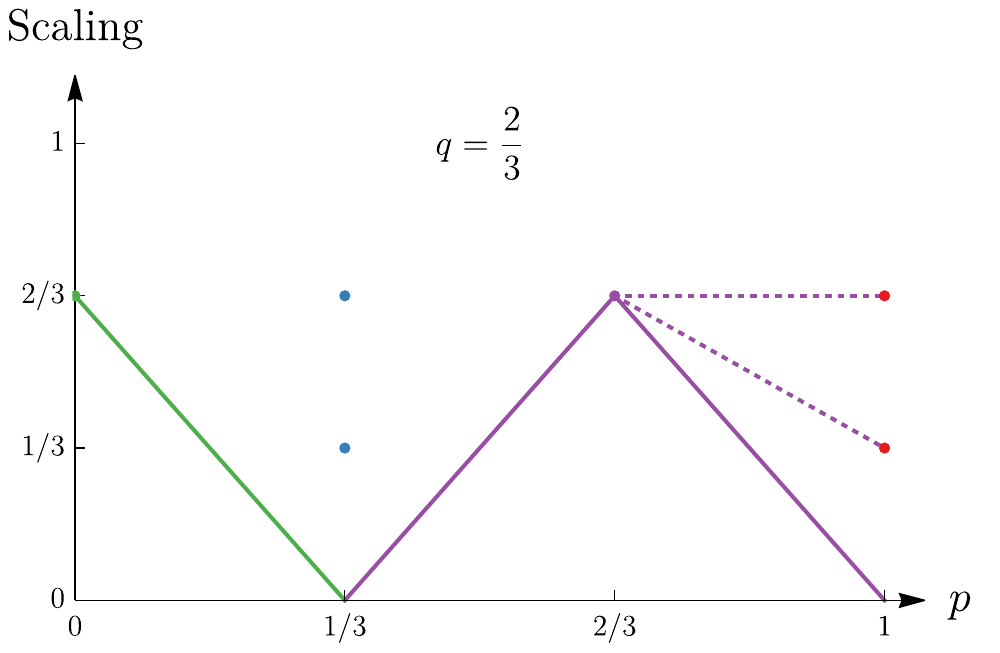}
	\caption{Top: critical curve in the backside-projected sky \eqref{eq:BacksideProjection} of a prograde ISCO emitter with $r_s=r_\mathrm{ms}^+=M(1+2^{1/3}\kappa^{2/3})$ orbiting a rapidly rotating BH with spin $a/M=99.8\%$ and deviation from extremality $\kappa\approx0.06$ (compare with Fig.~\ref{fig:BacksideProjection}).  The exact curve [Eq.~\eqref{eq:BacksideCriticalCurve}] is plotted in light gray, while the red, purple, and blue curves correspond to the subleading contributions from the NHEK $p$-bands with $p=1$, $q$, and $q/2$, respectively [Eqs.~\eqref{eq:p=1}--\eqref{eq:p=q/2}], and the green curve to the $p=0$ extreme Kerr subleading contribution [Eqs.~\eqref{eq:p=0}].  Bottom: these bands form the boundaries of different scaling regions where power laws break (Table~\ref{tbl:Scalings}), and gluing their segments together (plotted only in the upper half-plane for clarity) provides an excellent approximation to the exact curve.}
	\label{fig:SubleadingCriticalCurve}
\end{figure}

Thus far, we have considered the high-spin expansion \eqref{eq:DoubleLimit} only to leading order in the deviation from extremality $\kappa$. Previous works have rarely gone beyond leading order in this perturbative expansion. In this final section, we extend to subleading order our computation of the critical curve for an emitter in the throat.

In Sec.~\ref{sec:ExtremalCriticalCurve}, we found that at extremality, a double scaling limit---in which both the photon shell radius $\rt$ and orbiter radius $r_s$ are scaled into the throat---was required in order to resolve the full critical curve [Eq.~\eqref{eq:DoubleLimit}].  When the equatorial emitter is in the asymptotically flat region (i.e., $r_s=M(1+\kappa^0R_s)$ for finite nonzero $R_s$), the $p=0$ extreme Kerr portion of the photon shell gives the majority of the critical curve, with the $p=1$ NHEK band providing the remaining portion (the NHEKline).  The NHEK bands with scaling power $p\in\pa{0,1}$ only contribute the two points connecting these two parts.

On the other hand, in the sky of an equatorial orbiter in the throat, at some orbital radius outside the ISCO (i.e., $r_s=M(1+\epsilon^qR_s)$ for $0<q\le2/3$) but within the photon shell $\br{\rt_-,\rt_+}$, the critical curve splits into two parts $\rt>r_s$ and $\rt<r_s$ resolved by the $p=q/2$ and $p=1$ NHEK bands, respectively.  The NHEK bands with scaling power $p\in(q/2,1)$ resolve the two points connecting these parts, while those with scaling power $p\in(0,q/2)$ resolve the cusp on the backside-projected critical curve corresponding to the outer edge of the photon shell $r_+$.  Thus, different $p$-bands account for different parts of the critical curve, or at times, only a point.

For the latter case of a near-horizon emitter, it is interesting to consider the subleading correction in $\kappa$ to the critical curve.  This correction is computed by expanding Eq.~\eqref{eq:SkyCriticalCurve} beyond the $\mathcal{O}(\kappa^0)$ result \eqref{eq:CriticalCurveNHEK}.  Such an expansion yields several contributions with different power scalings in $\kappa$.  Which one of these contributions turns out to be most significant is determined by the specific values of $p$ and $q$.  The different possible scalings of subleading corrections to the critical direction cosines $\cos{\tilde{\Psi}}$ and $\cos{\tilde{\Theta}}$, which determine the curve, are shown in Table~\ref{tbl:Scalings}.

For finite deviation from extremality $0<\kappa\ll1$, the $\mathcal{O}(\kappa^0)$ approximation \eqref{eq:CplusNHEK}--\eqref{eq:CminusNHEK} is no longer close to the exact critical curve.  To improve it, one must include additional NHEK $p$-bands to cover intermediate parts of the curve, which the original bands no longer resolve.  The portions of the curve obtained from all these scaling regimes are to be glued to each other at intermediate points in order to produce a single curve that tracks the exact one (Fig.~\ref{fig:SubleadingCriticalCurve} top).  This suggests that, in order to approximate the critical curve to better and better precision as the near-extremal deviation $\kappa$ grows, one must include increasingly many NHEK $p$-bands with different scaling powers to fill in the gaps that appear at different positions along the curve.  Such a procedure would define a map in the emitter sky between position along the critical curve and different near-horizon scaling behaviors.

As a first step towards understanding this map, we note that for certain special values of $p$, the scaling exponent of the dominant subleading correction can change its functional form (Table~\ref{tbl:Scalings}), resulting in a transition in the scaling behavior (Fig.~\ref{fig:SubleadingCriticalCurve} bottom).  Empirically, we observe that it is precisely these $p$-bands where power laws break whose contributions must be included in order to obtain a good approximation of the exact curve.  This confirms that distinct NHEK bands really resolve different physics, and suggests that nonextremal BH physics may be recovered from a high-spin expansion as a series of successively broken power laws.  A more detailed study of this tantalizing possibility is deferred to future work.

To conclude, we explicitly describe how to obtain the colored segments approximating the critical curve of an emitter at the prograde ISCO radius $r_\mathrm{ms}^+$ (Fig.~\ref{fig:SubleadingCriticalCurve} top), which scales to the horizon with $q=2/3$.  In this case, the power laws describing the scaling of the corrections to the extremal direction cosines $\cos{\tilde{\Theta}}$ and $\cos{\tilde{\Psi}}$ are broken at $p=1/3$, $p=2/3$, and $1$ (Fig.~\ref{fig:SubleadingCriticalCurve} bottom).  For an approximation of the critical curve, we must then use these NHEK bands along with the boundary case: the $p=0$ extreme Kerr contribution. We expand the direction cosines to $\mathcal{O}(\kappa^{2/3})$ in each NHEK band, which is the lowest order for which $\cos{\tilde{\Psi}}$ and $\cos{\tilde\Theta}$ receive corrections to their extremal values in all four bands.

For $p=1$, we obtain
\begin{subequations}
\label{eq:p=1}
\begin{align}
    \cos{\tilde{\Psi}}&=\frac{2^{2/3}}{3\Rt}\kappa^{1/3}-\frac{A}{2^{2/3}}\kappa^{2/3},\\
    \cos{\tilde{\Theta}}&=\frac{B}{\Rt}\pa{1+\frac{2^{5/3}}{3\Rt}\kappa^{1/3}-2^{1/3}A\kappa^{2/3}},\\
    A=&1-\frac{8}{9\Rt^2},\quad
    B=\mp_\theta\sqrt{1-\frac{\Rt_-^2}{\Rt^2}}.
\end{align}
\end{subequations}
For $p=q$, we obtain
\begin{subequations}
\begin{align}
   \cos{\tilde{\Psi}}&=-\frac{2^{1/3}}{6}\pa{2^{1/3}\Rt^2-\frac{2^{4/3}}{\Rt}+3}\kappa^{2/3},\\
    \cos{\tilde{\Theta}}&=\mp_\theta\br{1-\frac{2^{1/3}}{3\Rt^2}\pa{2^{1/3}\Rt^2+1}\pa{\Rt-2^{1/3}}^2\kappa^{2/3}}.
\end{align}
\end{subequations}
For $p=q/2$, we obtain
\begin{subequations}
\label{eq:p=q/2}
\begin{align}
    \cos{\tilde{\Psi}}&=-\frac{\Rt^2}{Y}-\frac{3}{Y^2}\pa{2^{1/3}\Rt^4+\frac{9\Rt^2}{2^{4/3}}-3}\kappa^{2/3},\\ 
    \cos{\tilde{\Theta}}&=\mp_\theta\pa{\frac{3\times2^{1/3}}{Y}+\frac{2^{7/3}\Rt}{Y}\kappa^{1/3}}\notag\\
    &\mp_\theta\frac{2}{Y^2}\pa{\frac{2^{1/3}\Rt^4}{3}-\frac{7\Rt^2}{2^{4/3}}-9}\kappa^{2/3}\\
    Y&=2\Rt^2 +3\times2^{1/3}
\end{align}
\end{subequations}
For $p=0$, we obtain
\begin{subequations}
\label{eq:p=0}
\begin{align}
    \cos{\tilde{\Psi}}=&-\frac{1}{2}-\frac{3}{2^{5/3}}\pa{1-\frac{1}{\Rt^2}}\kappa^{2/3},\\
    \cos{\tilde{\Theta}}=&\mp_\theta\frac{\sqrt{3(3-\Rt)(1+\Rt)^3}}{2^{2/3}\Rt^2}\kappa^{2/3}.
\end{align}
\end{subequations}
Plugging these equations into the backside projection \eqref{eq:BacksideProjection} results in the approximation to the critical curve shown in the top panel of Fig.~\ref{fig:SubleadingCriticalCurve}.

\acknowledgements{We acknowledge Samuel Gralla, Achilleas Porfyriadis, and Andrew Strominger for useful conversations.  DG acknowledges support from NSF GRFP Grant No. DGE1144152.  SH and AL gratefully acknowledge support from the Jacob Goldfield Foundation.  Funding for shared facilities used in this research was provided by NSF Grant No. 1707938.}

\appendix

\section{Near-horizon geometry}
\label{app:NHEK}

In this appendix, we review the near-horizon geometry of a rapidly rotating BH, with spin $a=M\sqrt{1-\kappa^2}$ parametrized by its small deviation from extremality $0<\kappa\ll1$.  As discussed in Sec.~\ref{sec:ExtremalCriticalCurve}, the standard parametrization of the critical curve by photon shell radius $\rt$ breaks down in the extremal limit $\kappa\to0$.  As a result, we need to consider the near-horizon geometry separately from the far region covered by the extreme Kerr metric.

This breakdown occurs because a significant part of the photon shell penetrates the near-horizon region, which corresponds to a vanishingly small range of Boyer-Lindquist radii $(r-r_+)/r_+\ll 1$, where
\begin{align}
    r_+=M+\sqrt{M^2-a^2}
    \approx M(1+\kappa)
\end{align}
denotes the event horizon radius.  Nevertheless, the near-horizon geometry is known to be nondegenerate, as can be shown explicitly by introducing appropriate scaling coordinates adapted to the near-horizon physics \cite{Bardeen1999}.  This procedure yields the Near-Horizon Extreme Kerr (NHEK) metric.  Thus, in the extremal limit, the Kerr spacetime can be thought of as composed of two asymptotic regions---the NHEK throat and the asymptotically flat part---with the photon shell straddling the two.

\subsection{NHEK geometry and geodesics}

Starting out with the Kerr spacetime \eqref{eq:Kerr} in the high-spin regime $0<\kappa\ll1$, the geometry of the near-horizon region is resolved by introducing a one-parameter family of coordinate transformations \cite{Bardeen1999,Hadar2014},
\begin{align}
    t=\frac{2MT}{\kappa^p},\quad
    r=r_+\pa{1+\kappa^pR},\quad
    \phi=\Phi+\frac{T}{\kappa^p},
\end{align}
where $0<p\le1$ labels the so-called ``NHEK band'' \cite{Kapec2020}, and taking the $\kappa\to0$ limit.  Performing this procedure with $p<1$ gives the NHEK geometry,
\begin{gather}
	\frac{d\hat{s}^2}{2M^2\Gamma}=-R^2\ed T^2+\frac{\ed R^2}{R^2}+\ed\theta^2+\Lambda^2\pa{\ed\Phi+R\ed T}^2,\notag\\
	\label{eq:NHEK}
	\Gamma=\frac{1+\cos^2{\theta}}{2},\quad
	\Lambda=\frac{2\sin{\theta}}{1+\cos^2{\theta}}.
\end{gather}
Taking the $\kappa\to0$ limit with $p=1$, however, gives a different limiting metric known as near-NHEK \cite{Bredberg2010}, which depends explicitly on $\kappa$.  Remarkably, however, NHEK and near-NHEK are locally diffeomorphic.  In the astrophysical context, this fact can be exploited to analytically describe processes involving high-spin BHs \cite{Hadar2014}.

In NHEK, a particle of mass $\mu$ with affinely
parametrized trajectory $x^\nu(\tau)$ has four-momentum
\begin{align}
	P_\nu\ed X^\nu=&-\mathcal{E}\ed T\pm_R\frac{\sqrt{\mathcal{R}_n(R)}}{R^2}\ed R\notag\\
	&\quad\pm_\theta\sqrt{\Theta_n(\theta)}\ed\theta+\mathcal{L}\ed\Phi
\end{align}
given in terms of the radial and angular potentials
\begin{align}
	\mathcal{R}_n(R)&=\pa{\mathcal{E}+\mathcal{L}R}^2-\pa{\mathcal{C}+\mathcal{L}^2}R^2,\\
	\Theta_n(\theta)&=\mathcal{C}+\pa{1-\frac{1}{\Lambda^2}}\mathcal{L}^2-\pa{2M^2\Gamma}\mu^2.
\end{align}
The quantities $(\mu,\mathcal{E},\mathcal{L},\mathcal{C})$ are conserved along the trajectory, ensuring that NHEK geodesic motion is completely integrable.  The NHEK geodesic equation was only recently solved analytically in complete generality \cite{Kapec2020,Compere2020}.

\subsection{Circular equatorial orbiters}

The NHEK geometry admits (necessarily prograde) timelike circular equatorial geodesics at any orbital radius $R=R_s$, with conserved quantities obtained by solving $\Theta_n(\pi/2)=\Theta_n'(\pi/2)=0$ and $\mathcal{R}_n(R_s)=\mathcal{R}_n'(R_s)=0$, resulting in \cite{Hadar2014}
\begin{align}
	\mathcal{E}_s=\mathcal{C}_s=0,\quad
	\frac{\mathcal{L}_s}{\mu}=\frac{2M}{\sqrt{3}}.
\end{align}
These orbits necessarily have an angular velocity $\Omega_s$ and four-velocity $U_s^\nu$ given by
\begin{align}
	\Omega_s=-\frac{3}{4}R_s,\quad
	U_s=\frac{2}{\sqrt{3}MR_s}\pa{\pd_T+\Omega_s\pd_\Phi}.
\end{align}
The orbital motion is marginally stable, $\mathcal{R}_n''(R_s)=0$.  The local rest frame of the orbiter is conveniently described using an orthonormal tetrad with components aligned with the $(R,\theta)$ directions,
\begin{subequations}
\begin{align}
	\hat{\mathbf{e}}_{(T)}&=U_s,
	&&\hat{\mathbf{e}}_{(R)}=\frac{R}{M}\pd_R,\\
	\hat{\mathbf{e}}_{(\theta)}&=\frac{1}{M}\pd_\theta,
	&&\hat{\mathbf{e}}_{(\Phi)}=\frac{1}{\sqrt{3}MR_s}\pd_T,
\end{align}
\end{subequations}
obeying the conditions \eqref{eq:TetradConditions} everywhere along the orbit.

\subsection{Bound photon orbits}

The NHEK geometry also admits spherical bound photon orbits of fixed radius $R=\Rt$, with conserved quantities obtained by solving $\mathcal{R}_n(\Rt)=\mathcal{R}_n'(\Rt)=0$ to find
\begin{align}
    \label{eq:PhotonOrbitsNHEK}
    \tilde{\mathcal{E}}=\tilde{\mathcal{C}}=0,\quad
    \tilde{\mathcal{L}}>0,
\end{align}
with $\tilde{\mathcal{L}}$ otherwise arbitrary.  These bound photon orbits are also marginally stable because $\mathcal{R}_n''(\Rt)=0$.  Their polar motion oscillates around the equator between the critical angles $\theta_c=\arctan\pa{\sqrt{2}/3^{1/4}}$ and $\pi-\theta_c$, which satisfy $\Lambda(\theta_c)=1$ and correspond to hypersurfaces on which the NHEK metric is exactly AdS$_3$.

\subsection{Intrinsic NHEK critical curve}

Following Eqs.~\eqref{eq:PolarAngle}--\eqref{eq:ForwardCosine}, we define local angles of photon emission in the NHEK orbiter sky as
\begin{align}
    \cos{\Theta_n}&=-\frac{P^{(\theta)}}{P^{(T)}}
    =\mp_\theta\frac{\sqrt{\frac{4}{3}\mathcal{C}+\mathcal{L}^2}}{\frac{4}{3}\frac{\mathcal{E}}{R_s}+\mathcal{L}},\\
    \cos{\Psi_n}&=\frac{P^{(\phi)}}{P^{(T)}}
    =-\frac{1}{2+\frac{3}{2}\frac{\mathcal{L}}{\mathcal{E}}R_s}.
\end{align}
The bound photon orbits \eqref{eq:PhotonOrbitsNHEK} appear at angles
\begin{align}
    \cos{\Theta}_n=\mp_\theta1,\quad
    \cos{\Psi}_n=0,
\end{align}
in the sky, corresponding to emission towards the local zenith and nadir.  Physically, this result can be interpreted to mean that the NHEK geometry resolves a single photon shell radius, which happens to be precisely that where the orbiter lies.  This is consistent with the discussion about the location of the extremal ISCO in Ref.~\cite{Gralla2015}: since the NHEK geometry \eqref{eq:NHEK} is invariant under rescalings $(T,R)\to(T/\lambda,\lambda R)$, all the radii can be mapped into each other under dilations and are therefore in some sense identical; it is only by reattaching the asymptotically flat region that the scale-invariance is broken and differences in NHEK radius re-acquire physical meaning.

As such, an intrinsic NHEK calculation is not sufficient to resolve the critical curve of a NHEK emitter.  Its determination truly requires matching across the different asymptotic regions of the high-spin BH, possibly using more than one NHEK band as in Sec.~\ref{sec:PerturbationTheory}.

\bibliographystyle{utphys}
\bibliography{KerrFlux.bib}

\end{document}